\documentclass[draftcls,onecolumn,12pt]{IEEEtran}

\usepackage{threeparttable}
\usepackage{amsthm}
\usepackage{subfigure}
\usepackage{cite}

\ifCLASSINFOpdf
\else
   \usepackage[dvips]{graphicx}
   \graphicspath{{../eps/}}
   \DeclareGraphicsExtensions{.eps}
\fi

\usepackage[cmex10]{amsmath}
\usepackage{empheq}

\usepackage{mdwmath}
\usepackage{mdwtab}
\begin{document}
%\title{Relay Placement in Limited Feasible Regions for Multi-hop FSO System Over Lognormal Channel When Obstacles Exist}
\title{A New Asymptotic Analysis Technique for Diversity Receptions Over Correlated Lognormal Fading Channels}

\author{
\IEEEauthorblockN{Bingcheng Zhu\IEEEauthorrefmark{1}, Julian Cheng\IEEEauthorrefmark{2},~\IEEEmembership{Senior Member,~IEEE}, Jun Yan\IEEEauthorrefmark{1}, Jinyuan Wang\IEEEauthorrefmark{1}, Lenan Wu\IEEEauthorrefmark{3} and Yongjin Wang\IEEEauthorrefmark{1}}\\
\thanks{Bingcheng Zhu, Jun Yan, Jinyuan Wang and Wang Yongjin are with Gr\"{u}nberg Research Centre, Nanjing University of Posts and Telecommunications, Nanjing, China, (e-mails: \{zbc,yanj,jywang,wangyj\}@njupt.edu.cn). Julian Cheng is with the School of Engineering, The University of British Columbia, Kelowna, BC, Canada, (e-mail: julian.cheng@ubc.ca). Lenan Wu is with the School of Information Science and Engineering, Southeast University,  Nanjing, China (e-mail: wuln@seu.edu.cn). This work is supported by National Science Foundation of China (61322112, 61531166004); NUPTSF(NY216008); Young Elite Scientist Sponsorship Program by CAST.}
}
\maketitle
\thispagestyle{empty}

\begin{abstract}
Prior asymptotic performance analyses are based on the series expansion of the moment-generating function (MGF) or the probability density function (PDF) of channel coefficients. However, these techniques fail for lognormal fading channels because the Taylor series of the PDF of a lognormal random variable is zero at the origin and the MGF does not have an explicit form. Although lognormal fading model has been widely applied in wireless communications and free-space optical communications, few analytical tools are available to provide elegant performance expressions for correlated lognormal channels. In this work, we propose a novel framework to analyze the asymptotic outage probabilities of selection combining (SC), equal-gain combining (EGC) and maximum-ratio combining (MRC) over equally correlated lognormal fading channels. Based on these closed-form results, we reveal the followings: i)  the outage probability of EGC or MRC becomes an infinitely small quantity compared to that of SC at large signal-to-noise ratio (SNR); ii) channel correlation can result in an infinite performance loss at large SNR. More importantly, the analyses reveal insights into the long-standing problem of performance analyses over correlated lognormal channels at high SNR, and circumvent the time-consuming Monte Carlo simulation and numerical integration.
\end{abstract}

%\def\IEEEkeywordsname{Keywords}

% \fi
\IEEEpeerreviewmaketitle

\section{Introduction}
Diversity reception systems combine signals suffering different channel fading in order to obtain a more reliable output signal\cite{goldsmith2005wireless}. The simplest diversity reception scheme is selection combining (SC) which selects the channel with the highest signal-to-noise ratio (SNR). Maximum-ratio combining (MRC) is the optimal linear diversity reception technique that combines all of the channels with the optimal weights, but such operation requires phase and fading amplitude information of the channels. Equal-gain combining (EGC) combines the channels with equal weights, and it usually provides performance close to MRC without requiring the amplitude information.

Exact performance analyses of diversity receptions over Rayleigh, Rician and Nakagami-$m$ fading channels are relatively straightforward. Closed-form or single-fold-integral outage probability expressions have been derived for independent channels \cite[Chaps. 6, 9]{simon2005digital}\cite[Chap. 7]{goldsmith2005wireless}. Existing works have also provided closed-form or single-fold-integral expressions for error rates \cite{MallikHSMRC,Win_Nakagami_MRC,Win_Virtual_Rayleigh,McKay_MIMO_MRC_Rayleigh} and outage probabilities \cite{Beaulieu_GeneralizedCorrelation,McKay_MIMO_MRC_Rayleigh,Weibull_theory} for diversity receptions over correlated channels. In contrast, performance
analysis of diversity receptions over correlated lognormal fading channels is much more challenging, which leads to $(L-1)$-fold nested integrals \cite{DiversityLognormal2} where $L$ is the number of links, and these integrals are troublesome to be estimated using numerical methods. The time complexity of numerical integration increases exponentially with the number of channels, thus it is not practical to perform numerical integration when the number of receptions becomes large. Even for the dual-branch cases, the exact outage probability expressions of SC over correlated lognormal fading channels can only be simplified to a single-fold integral \cite{Alouini_lognormalDiversity}.

To circumvent the difficulty of the exact performance analyses and numerical estimation, various approximation techniques have been proposed for the diversity systems over lognormal fading channels. In \cite{schwartz1982distribution,fenton1960sum,Pratesi_MomentsMatchingLognormal,Log_shifted_gamma,Sum_lognormal_LSN}, the authors approximated the probability density function (PDF) of sum of lognormal random variables (RVs) using another lognormal RV by matching their moments, and these techniques were widely applied in subsequent studies due to its simplicity \cite{RenzoCooperative_lognormal,safari2008relay,navidpour2007ber}. In \cite{OutageLognormalVictor,PhibooOutageSC,Mehta_sumLognormal}, the authors applied the Gaussian-Hermite integration technique to numerically estimate the outage probability of MRC over lognormal fading channels. In \cite{Beaulieu_sumlognormal,Beaulieu_optimal_sumOflognormal,QuadratureBasedLognormalSumApproximation}, the authors approximated the cumulative distribution function (CDF) of sum of lognormal in a transformed domain. In \cite{BeaulieuCompareSumLognormal}, various ways of approximating the sum of lognormal RVs are compared. However, all of the aforementioned approximation methods cannot provide reliable estimation in high SNR region. Bounds on the CDF of sum of lognormal RVs were studied in \cite{BoundsLognormal,Alouini_lognormalDiversity}, but these bounds cannot provide accurate outage probability estimation at large SNR either. Some works rely on more complicated random variables to approximate the sum of lognormal RVs, and determine the associated parameters using numerical methods \cite{Sum_lognormal_Power_lognormal,SLN_Pearson,MixtureLognormal_SLN}. However, these approximation techniques suffer larger time complexity and reveal few insights.

Asymptotic analysis is a kind of approximations that can provide accurate performance estimation in large SNR region. Over Rayleigh, Rician, Nakagami-$m$ and most other fading channels, closed-form asymptotic error rate and outage probability expressions of MRC, EGC and SC have been obtained with arbitrary correlation structure \cite{wang2003simple,BingchengNakagami,BingchengHierarchical,XueguiDiversity}. Asymptotic outage probability expressions have also been derived for free-space optical (FSO) communications following Gamma-Gamma fading channels \cite{Fan,GammaGammaMIMO}. Unfortunately, the classical asymptotic analysis techniques in \cite{wang2003simple,BingchengNakagami,BingchengHierarchical,XueguiDiversity,Fan,GammaGammaMIMO} fail to provide meaningful result due to the followings:
\begin{itemize}
  \item Moment-generating function (MGF) of lognormal PDF does not have a unified explicit expression\cite{Alouini_lognormalDiversity}. Therefore, all methods based on MGF fail to work with the lognormal channels.
  \item The Taylor series of the PDF of a lognormal RV is zero at the origin, and this results in an infinite diversity order \cite{DiversityLognormal}. This implies that the asymptotic outage probability is zero, which is a meaningless result because it cannot quantify the performance gap between two systems with different branch number and correlation status.
  \item CDF of a sum of lognormal RVs does not have a closed-form expression and it is challenging to be accurately approximated at the origin.
\end{itemize}
Due to the above difficulties, few works studied the asymptotic outage probabilities of diversity systems over lognormal fading channels. For EGC and MRC, the problem is equivalent to the CDF left tail approximation of sum of lognormal RVs, and for SC the problem can be reduced to the asymptotic approximation of multi-variate lognormal CDF. In \cite{szyszkowicz2007tails}, the authors used the CDF tail of another lognormal RV to approximate the CDF tail of sum of independent lognormal RVs, but it was subsequently proved in \cite{AsymptoticSumOfTwoLognormal} that ``any lognormal, reciprocal Gamma or log shifted Gamma cannot be used to fit the left tail, under the independence hypothesis''. The authors in \cite{AsymptoticSumOfTwoLognormal} derived the approximation of the left tail of the PDF of sum of two correlated lognormal RVs, but asymptotic CDF expression was not derived. In \cite{gulisashvili2016tail}, the approximate CDF of the sum of lognormal RVs was transformed into a quadratic optimization problem, which relies on recursive algorithms. For lognormal fading channels, asymptotic performance expression is both theoretically and practically important. This is because the diversity order of lognormal fading channels is infinite \cite{DiversityLognormal}, which results in a dramatic decrease in outage probability as SNR increases, thus it can be unacceptably time-consuming to estimate the performance of diversity systems using Monte Carlo simulation in large SNR region because it requires many channel coefficient samples to reliably estimate the ultra-low outage probability ($<10^{-12}$).

Since it is challenging to perform exact analyses, accurate approximations and asymptotic analyses for correlated lognormal channels, much fewer insights have been revealed for FSO multiple-input multiple-output (MIMO) links suffering weak turbulence-induced fading \cite{AtmosphericTurbulence} and wireless MIMO links suffering slow fading \cite{Alouini_lognormalDiversity,DiversityLognormal2}, which hampers the system design.

In this work, a new theorem is developed to simplify the asymptotic analyses of lognormal fading channels. Based on this theorem, we derive closed-form asymptotic outage probabilities of SC, EGC and MRC over equally correlated lognormal fading channels. For SC, the derived asymptotic outage probability is expressed using elementary functions. For EGC and MRC,  the derived asymptotic outage probabilities are expressed using Marcum-$Q$ functions. Two properties of lognormal fading channels are revealed: i) the outage probability of EGC or MRC is an infinitely small quantity compared to the outage probability of SC as SNR approaches infinity where the channel correlation coefficients are fixed. ii) Channel correlation will induce infinite SNR loss at high SNR. Both properties are in sharp contrast with the other fading channels (Rayleigh, Rician, Nakagami-$m$, Gamma-Gamma, etc.). Compared to the methods in prior works \cite{AsymptoticSumOfTwoLognormal,gulisashvili2016tail}, the derivation in this work is much simpler and has an elegant geometrical interpretation. Numerical results show that the proposed asymptotic expressions are highly accurate in medium to high SNR region. More importantly, new insights into the long-standing problem are revealed, and one can efficiently evaluate the performance of a diversity system over lognormal fading channels without resorting to the expensive Monte Carlo simulation or numerical integration.

The remainder of this paper is organized as follows. Section \ref{SystemModel} introduces the system model including the channel model, the correlation model and the outage probability of diversity receptions. In Section \ref{TheoremSection}, we propose a key theorem of the integrals of joint Gaussian PDF, and the theorem is applied in Section \ref{SCoutage} to derive the asymptotic outage probability expressions for SC. The asymptotic outage probabilities of EGC and MRC are derived in Section \ref{EGCMRCoutage}. In Section \ref{discussions}, we discuss the essential differences between lognormal fading channels and the other fading channels, and compare the outage probabilities of SC, EGC and MRC. Numerical results are presented in Section \ref{numerical}, and Section \ref{conclusions} draws some conclusions.

\section{System Model}\label{SystemModel}

\subsection{Correlated Lognormal Fading Channels}
We assume that the channel coefficients of the $L$ links are $\textbf{c}=\left[e^{G_1},\cdots,e^{G_L}\right]^T$, where $[G_1,\cdots,G_L]^T$ is a correlated Gaussian random vector, whose elements have identical mean $\mu_G$ and variance $\sigma_G^2$, and $[\cdot]^T$ denotes the transpose operation. The equality between statistics of links is valid in most FSO and wireless MIMO links. $\sigma_G$ is known as the ``dB spread'' in mobile radio environment and $\sigma_G^2$ is proportional to the Rytov variance in FSO communications. The received signal vector is
\begin{equation}\label{II-0.5}
  \textbf{y}= \textbf{c} x+\textbf{n}
\end{equation}
where $x$ is a real-value signal;  $\textbf{n}$ is an $L\times1$ Gaussian random vector denoting the additive white Gaussian noise and we assume $E[\textbf{n}\textbf{n}^T]=\textbf{I}_L$ without loss of generality, where $\textbf{I}_L$ is an $L\times L$ identity matrix, and $E[\cdot]$ denotes the expectation.

After photoelectric conversion, the average received electrical power of the channels can be calculated as ${{\bar E}_r}=E[e^{2G_l}]=e^{2\mu_G+2\sigma_G^2},\forall l=1,\cdots,L$. The $L$ correlated Gaussian RVs $G_l$'s are generated by $L$ independent Gaussian RVs $X_l$'s with the following relationship
\begin{equation}\label{II-1}
\left\{ \begin{array}{l}
{G_1} = a{X_1} + {X_2} +  \cdots  + {X_L}\\
{G_2} = {X_1} + a{X_2} +  \cdots  + {X_L}\\
 \vdots \\
{G_L} = {X_1} + {X_2} +  \cdots  + a{X_L}
\end{array} \right.
\end{equation}
where $X_l\sim {\cal{N}} (\mu_X,\sigma_X^2),\forall l=1,\cdots,L$, and $a\in\left[1,\infty\right)$ is a parameter determining the correlation coefficients of $G_l$'s. According to \eqref{II-1}, the parameters between $X_l$'s and $G_l$'s must have the relationship ${\mu _X} = \frac{{{\mu _G}}}{{a + L - 1}}$ and
$\sigma _X^2 = \frac{{\sigma _G^2}}{{{a^2} + L - 1}}$.

According to \eqref{II-1}, the correlation coefficient between $G_m$ and $G_n$ ($m\neq n$) can be calculated as
\begin{equation}\label{II-4}
\begin{split}
{\rho} &= \frac{{E\left[ {\left( {{G_n} - {\mu _G}} \right)\left( {{G_m} - {\mu _G}} \right)} \right]}}{{\sigma _G^2}}=\frac{2a +L - 2}{{{a^2} +  {L - 1}}}.
\end{split}
\end{equation}
When $a\to\infty$ we have $\rho\to0$ which corresponds to the independent channels, and when $a=1$ we have $\rho=1$ which implies that the channels are identical. There must exist at least one real solution of $a$ given a fixed $\rho$ as
\begin{equation}\label{II-4.5}
  a = \frac{{1 + \sqrt {1 - \rho \left( {\rho \left( {L - 1} \right) - L + 2} \right)} }}{\rho }
\end{equation}
because the discriminant of \eqref{II-4} is $4\left( {1 - \rho } \right) + 4\rho \left( {1 - \rho } \right)\left( {L - 1} \right)>0$ for $\rho\in[0,1)$.

%In an FSO system, the known parameters include the average link optical power ${{\bar E}_r}=e^{\mu_G+\sigma_G^2/2}$, the correlation coefficient $\rho$ and the Rytov variance $\sigma_G^2$; $\mu_G$ can be solved based on the known values ${{\bar E}_r}$ and $\sigma_G^2$. $\mu_X$ and $\sigma_X^2$ can be calculated according to \eqref{II-2} and \eqref{II-3}, and the parameter $a$ can be calculated by solving the quadratic equation in \eqref{II-4}.

\subsection{Diversity Receptions Over The Correlated Lognormal Fading Channels}
SC selects the channel with the highest fading amplitude, and the instantaneous output SNR can be expressed as
\begin{equation}\label{II-5}
  \gamma_{SC}=\max_{l} \left\{e^{2G_l}\right\}.
\end{equation}
Outage occurs when $\gamma_{SC}$ falls below a predetermined threshold $\gamma_{th}$. The outage probability of SC can be calculated as
\begin{equation}\label{II-6}
  \begin{split}
{P_{out}^{SC}}\left( {{\gamma _{th}}} \right) &= \Pr \left\{ {\max_l \left\{ \exp(G_l) \right\} \le \sqrt{\gamma _{th}}} \right\}= \Pr \left\{ {\max_l \left\{ {\exp \left( {a{X_l} + \sum\limits_{k=1,k \ne l}^L {{X_k}} } \right)} \right\} \le \sqrt{\gamma _{th}}} \right\}.
\end{split}
\end{equation}

EGC combines the $L$ branches with equal weights, and the output SNR is
\begin{equation}\label{II-6.1}
  {\gamma _{EGC}} = \frac{1}{L}{\left( {\sum\limits_{l = 1}^L {{e^{G_l}}} } \right)^2}
\end{equation}
and the outage probability of EGC is
\begin{equation}\label{II-6.2}
\begin{split}
  P_{out}^{EGC}\left(\gamma _{th}\right)&=\Pr\left(\sum\limits_{l = 1}^L {\exp \left( G_l \right)}  \le \sqrt {L{\gamma _{th}}} \right)=\Pr\left(\sum\limits_{l = 1}^L {\exp \left( {a{X_l} + \sum\limits_{k=1,k \ne l}^L {{X_k}} } \right)}  \le \sqrt {L{\gamma _{th}}} \right).
  \end{split}
\end{equation}

MRC combines the $L$ branches with the optimal weights, and the output SNR is
\begin{equation}\label{II-6.3}
  {\gamma _{MRC}}{\rm{ = }}\sum\limits_{l = 1}^L e^{2G_l}
\end{equation}
and the outage probability of MRC is
\begin{equation}\label{II-6.4}
\begin{split}
  P_{out}^{MRC}\left(\gamma _{th}\right)&=\Pr\left(\sum\limits_{l = 1}^L {\exp \left( {2G_l} \right)}  \le {\gamma _{th}}\right)=\Pr\left(\sum\limits_{l = 1}^L {\exp \left( {2\left( {a{X_l} + \sum\limits_{k=1,k \ne l}^L {{X_k}} } \right)} \right)}  \le {\gamma _{th}}\right).
  \end{split}
\end{equation}

\section{A Useful Theorem For The Asymptotic Analysis}\label{TheoremSection}
\textbf{Lemma}: For the joint PDF of $L$ independent Gaussian RVs
\begin{equation}\label{III-1}
  {f_{\textbf{x},iid}}\left( {\textbf{x}} \right) = \frac{1}{{\sqrt {{{\left( {2\pi } \right)}^L}{\sigma ^{2L}}} }}\exp \left( { - \frac{1}{{2{\sigma ^2}}}{{\left| {{\textbf{x}} - {\boldsymbol{\mu}}} \right|^2}}} \right)
\end{equation}
where $\textbf{x}=[x_1,\cdots,x_L]^T$, and $\boldsymbol{\mu}=[\mu_1,\cdots,\mu_L]^T$, and $\sigma^2$ is the variance, and $|\cdot|$ denotes the 2-norm of a vector, the following equation holds
\begin{equation}\label{III-2}
  \lim_{|\boldsymbol{\mu}|\to\infty}\frac{{\int\limits_{{\Omega _1(\boldsymbol{\mu},\textbf{x}_0)}} {{f_{{\textbf{x}},iid}}\left( {\textbf{x}} \right)d} {\textbf{x}}}}{{\int\limits_{{\omega _2(\textbf{x}_0)}} {{f_{{\textbf{x}},iid}}\left( {\textbf{x}} \right)d} {\textbf{x}}}}=0
\end{equation}
where $d\textbf{x}\buildrel \Delta \over =dx_1\cdots dx_L$ and
\begin{equation}\label{III-2.1}
  {\Omega _1}(\boldsymbol{\mu},\textbf{x}_0) \buildrel \Delta \over = \left\{ {{\textbf{x}}\Big| {{{{\left| \textbf{x} - \boldsymbol{\mu} \right|}}}  > \left| {{{\textbf{x}}_0} - \boldsymbol{\mu}} \right| + \sqrt L \varepsilon+\varepsilon} } \right\}
\end{equation}
where $\textbf{x}_0=[x_{1,0},\cdots,x_{L,0}]^T$ and
\begin{equation}\label{III-2.2}
  {\omega _2}(\textbf{x}_0) \buildrel \Delta \over = \left\{ {\textbf{x}}\Big | {\left| {{x_l} - {x_{l,0}}} \right| \le \varepsilon ,\forall l = 1, \cdots ,L}  \right\}
\end{equation}
where $\varepsilon$ is an arbitrarily small positive constant.

\textbf{Proof:} see Appendix \ref{Lemma}.

\begin{figure}
  \centering
  \includegraphics[width=0.5\linewidth]{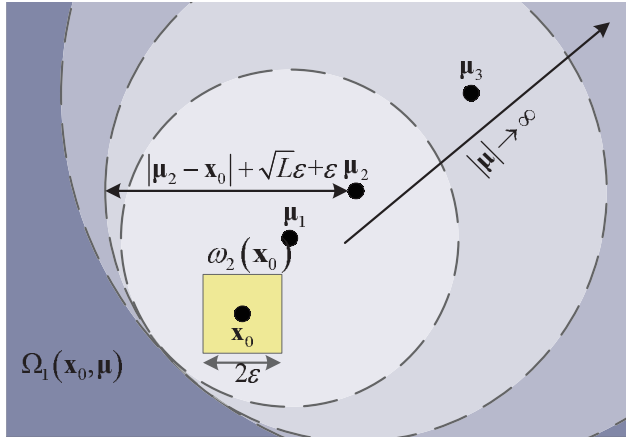}\\
  \caption{A diagram illustrating the relationship between the regions $\omega_2(\textbf{x}_0)$ and $\Omega_1(\textbf{x}_0)$, and points $\textbf{x}_0$ and $\boldsymbol{\mu}$ in the lemma on a two-dimensional plane, where $\left|\boldsymbol{\mu}_1\right|<\left|\boldsymbol{\mu}_2\right|<\left|\boldsymbol{\mu}_3\right|$. }\label{TheoremIllustration}
\end{figure}

The lemma in \eqref{III-2} essentially states that if the multi-variate independent Gaussian PDF is the integrand, the integral in an arbitrarily small hypercube ${\omega _2}(\textbf{x}_0)$ is a high order infinitely large quantity compared to the integral outside the hyperspherical region centered at $\boldsymbol{\mu}$ with radius $\left| {{{\textbf{x}}_0} - \boldsymbol{\mu}} \right| + \sqrt L \varepsilon+\varepsilon$. Figure \ref{TheoremIllustration} illustrates ${\Omega _1}(\boldsymbol{\mu},\textbf{x}_0)$ and ${\omega _2}(\textbf{x}_0)$ on a two-dimensional plane.

\textbf{Theorem}: For $L$-variate independent Gaussian PDF in \eqref{III-1},
%\begin{equation}\label{III-11}
%  {f_{\textbf{x},iid}}\left( {\textbf{x}} \right) = \frac{1}{{\sqrt {{{\left( {2\pi } \right)}^L}{\sigma ^{2L}}} }}\exp \left( { - \frac{1}{{2{\sigma ^2}}}{{\left| {{\textbf{x}} - {\boldsymbol{\mu}}} \right|^2}}} \right)
%\end{equation}
the following equation holds
\begin{equation}\label{III-12}
  \lim_{|\boldsymbol{\mu}|\to\infty}\frac{{\int\limits_{{\omega _1(\boldsymbol{\mu},\textbf{x}_0)}} {{f_{{\textbf{x}},iid}}\left( {\textbf{x}} \right)d} {\textbf{x}}}}{{\int\limits_{{\Omega _2(\textbf{x}_0)}} {{f_{{\textbf{x}},iid}}\left( {\textbf{x}} \right)d} {\textbf{x}}}}=0
\end{equation}
where $\omega _1(\boldsymbol{\mu},\textbf{x}_0)$ is any region with non-zero volume\footnote{The volume of an integral region $f_\textbf{x}(\textbf{x})<0$ is $\int\limits_{f_\textbf{x}(\textbf{x})<0}d\textbf{x}$.} contained by
\begin{equation}\label{III-13}
  \bar \Theta \left( {{{\textbf{x}}_0}}, \boldsymbol{\mu} \right) \buildrel \Delta \over = \left\{ {{\textbf{x}}\Big| {\left| {{\textbf{x}} - \boldsymbol{\mu}} \right| > \left| {{{\textbf{x}}_0} - \boldsymbol{\mu}} \right|} } \right\}
\end{equation}
and $\Omega _2(\textbf{x}_0,\boldsymbol{\mu})$ is any region with non-zero volume contained by
\begin{equation}\label{III-14}
\Theta \left( {{{\textbf{x}}_0}}, \boldsymbol{\mu} \right) \buildrel \Delta \over = \left\{ {{\textbf{x}}\Big| {\left| {{\textbf{x}} - \boldsymbol{\mu}} \right| < \left| {{{\textbf{x}}_0} - \boldsymbol{\mu}} \right|} } \right\}.
\end{equation}
\textbf{Proof:} see Appendix \ref{Theorem}.

\begin{figure}
  \centering
  \includegraphics[width=0.5\textwidth]{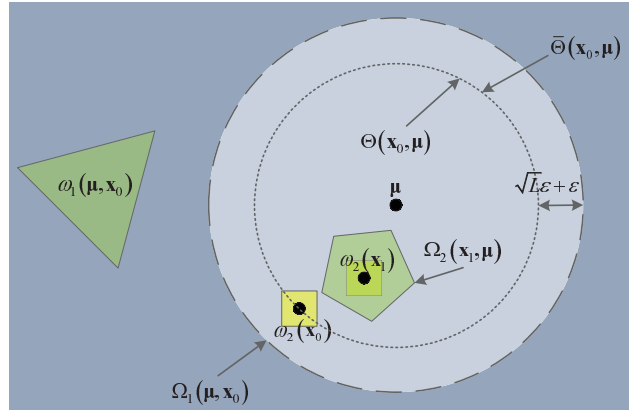}\\
  \caption{A diagram illustrating the relationship between the sets ($\bar \Theta \left( {{{\textbf{x}}_0}}, \boldsymbol{\mu} \right)$, $ \Theta \left( {{{\textbf{x}}_0}}, \boldsymbol{\mu} \right)$, $\Omega_2(\textbf{x}_1,\boldsymbol{\mu})$, $w_1(\boldsymbol{\mu}, \textbf{x}_0)$, $w_2(\textbf{x}_0)$ and $w_2(\textbf{x}_1)$) and points ($\textbf{x}_0$, $\textbf{x}_1$ and $\boldsymbol{\mu}$) in the lemma on a two-dimensional plane.}\label{TheoremFig}
\end{figure}

\begin{figure}
  \centering
  \subfigure[]{
    \label{fig:subfig:a} %% label for first subfigure
    \includegraphics[width=0.31\textwidth]{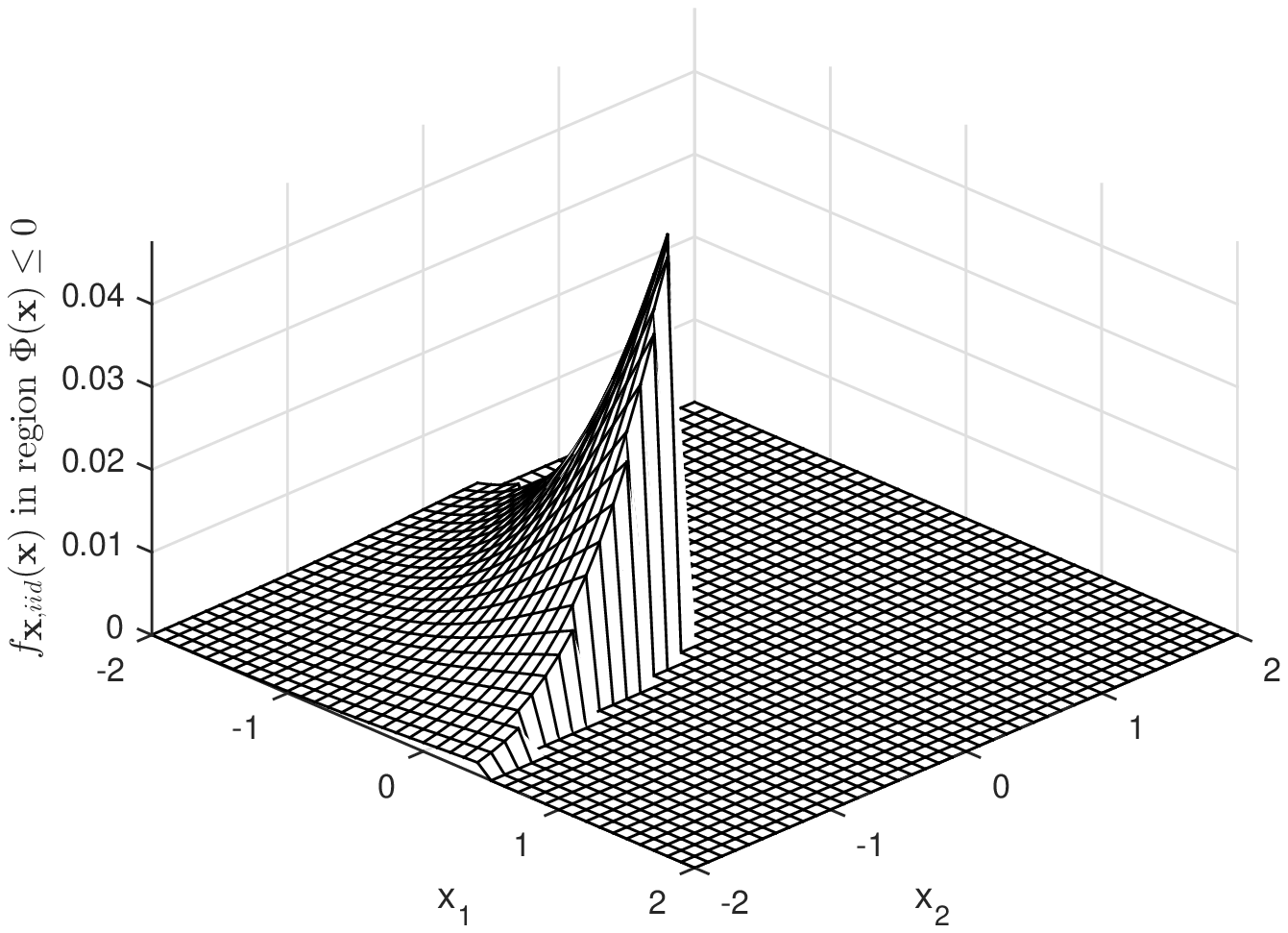}}
  \subfigure[]{
    \label{fig:subfig:b} %% label for second subfigure
    \includegraphics[width=0.31\textwidth]{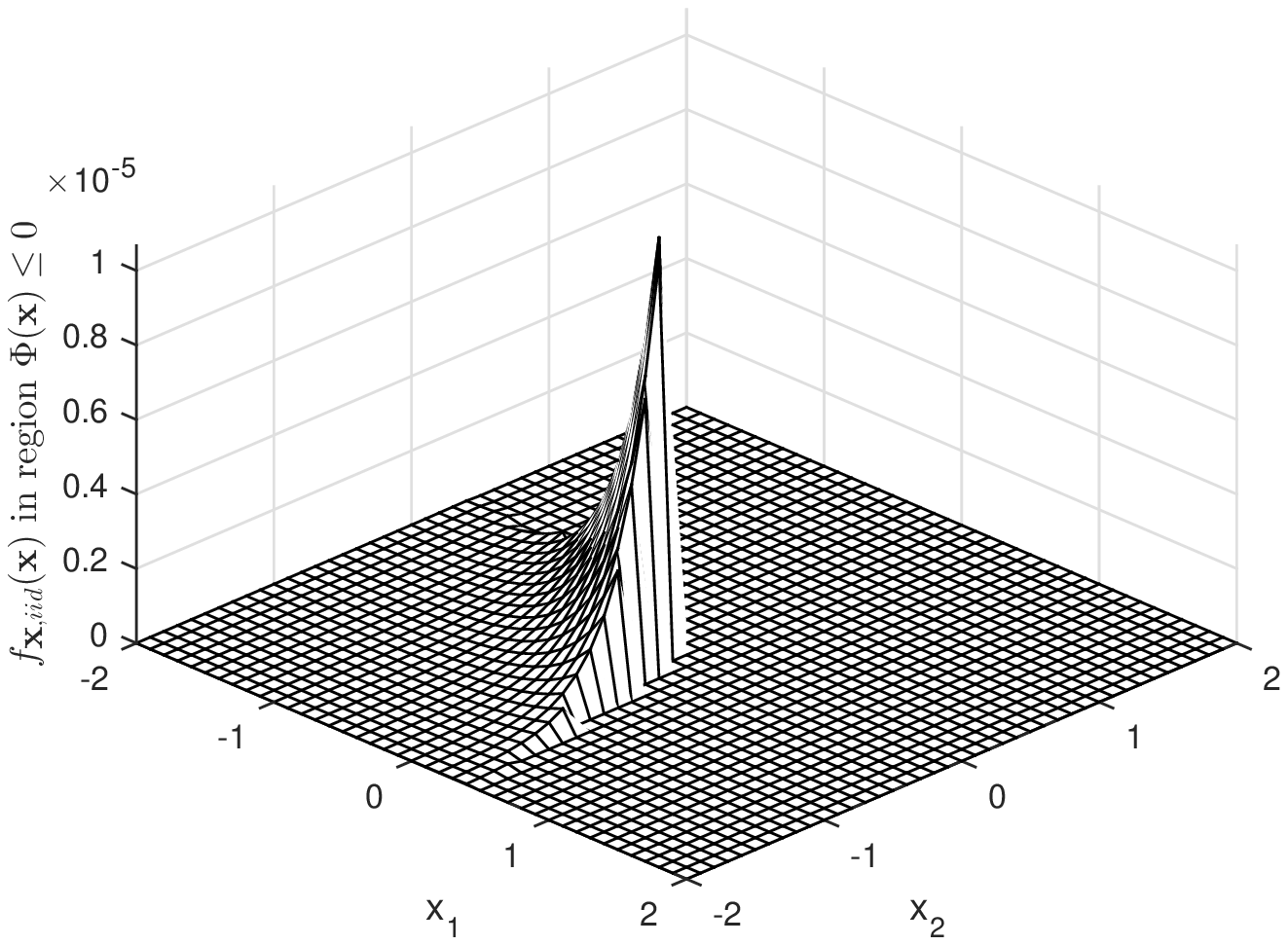}}
  \subfigure[]{
    \label{fig:subfig:c} %% label for second subfigure
    \includegraphics[width=0.31\textwidth]{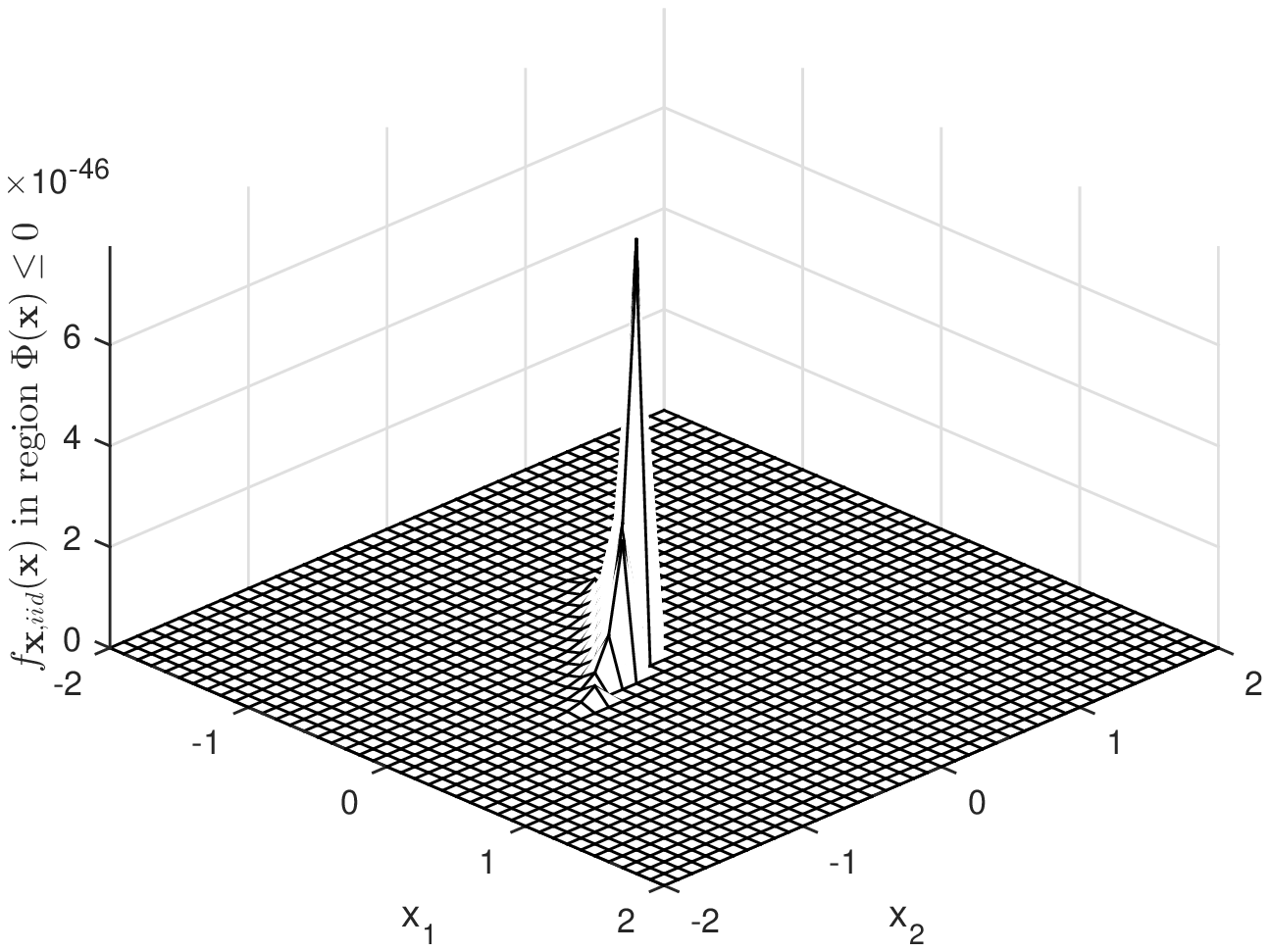}}
  \caption{Joint PDF $f_{\textbf{x},iid}(\textbf{x})$ in a region $\Phi(\textbf{x})\leq0$ where $\Phi \left( {\textbf{x}} \right) \buildrel \Delta \over = \exp \left( {a{x_1} + {x_2}} \right) + \exp \left( {{x_1} + a{x_2}} \right) - \sqrt {2{\gamma _{th}}} $. $L=2$, $a=5$, $\sigma=1$, $\gamma_{th}=1$. (a) $\boldsymbol{\mu}=[1,1]^T$. (b) $\boldsymbol{\mu}=[3,3]^T$. (c) $\boldsymbol{\mu}=[10,10]^T$.}
  \label{subfigs} %% label for entire figure
\end{figure}

The theorem in \eqref{III-12} says that the integral over any region within the hypersphere $\Theta \left( {{{\textbf{x}}_0}}, \boldsymbol{\mu} \right)$ is always an infinitely large quantity compared to the integral over any region outside $\Theta \left( {{{\textbf{x}}_0}}, \boldsymbol{\mu} \right)$ when $|\boldsymbol{\mu}|\to\infty$. Figure \ref{TheoremFig} illustrates the relationship between the involved regions in \eqref{III-12} on a two-dimensional plane. As a result, to obtain the asymptotic value of an integral with Gaussian integrand, it is valid to approximate the original integral region with its arbitrary subset as long as the approximating region keeps the dominant term. As an intuitive example, Fig. \ref{subfigs} presents $f_{\textbf{x},iid}(\textbf{x})$ in a region $\Phi(\textbf{x})\leq0$ on a two-dimensional plane with various mean vectors, where $\Phi \left( {\textbf{x}} \right) \buildrel \Delta \over = \exp \left( {a{x_1} + {x_2}} \right) + \exp \left( {{x_1} + a{x_2}} \right) - \sqrt {2{\gamma _{th}}}$. It can be observed that the region containing dominant PDF values becomes smaller and smaller as $\left|\boldsymbol{\mu}\right|$ grows.

\section{Asymptotic Outage Probability of SC Over Correlated Channels}\label{SCoutage}
Based on \eqref{II-6} and \eqref{III-1}, by letting $\sigma=\sigma_X$ and $\boldsymbol{\mu}=\boldsymbol{\mu}_X=[\mu_X,\cdots,\mu_X]^T$, we obtain
\begin{equation}\label{III-25}
  \begin{split}
{P_{out}^{SC}}\left( {{\gamma _{th}}} \right) &=\int\limits_{{\mathop {\max }\limits_l } \left\{ {\exp \left( {a{x_l} + \sum\limits_{k \ne l} {{x_k}} } \right)} \right\} < \sqrt{\gamma _{th}}} {{f_{{\textbf{x}},iid}}\left( {\textbf{x}} \right)} d{\textbf{x}}\\
 &= \int\limits_{a{x_l} + \sum\limits_{k \ne l} {{x_k}}  < \ln \sqrt{\gamma _{th}},\forall l=1,\cdots,L} {\frac{1}{{\sqrt {{{\left( {2\pi } \right)}^L}\sigma _X^{2L}} }}\exp \left( { - \frac{1}{{2\sigma _X^2}}|\textbf{x}-\boldsymbol{\mu}_X|^2 } \right)} d{\textbf{x}}
\end{split}
\end{equation}
and the integral region can be denoted as
\begin{equation}\label{III-25.5}
\Phi_{SC}(\textbf{x}) \buildrel \Delta \over = \mathop {\max }\limits_l \left\{ {\exp \left( {a{x_l} + \sum\limits_{k \ne l} {{x_k}} } \right)} \right\} - \sqrt {{\gamma _{th}}} \leq0.
\end{equation}
Unfortunately, it is challenging to simplify \eqref{III-25} further due to the nested integral region. However, it can be implied by the theorem in \eqref{III-12} that it is valid to approximate the integral region in \eqref{III-25} with its subset as long as the subset contains the dominant term. A necessary condition for the approximating subset to contain the dominant term is that it must contain a continuous set that contains or touches the nearest point to $\boldsymbol{\mu}_X$ in $\Phi_{SC}(\textbf{x})$, which can be proved by the theorem in \eqref{III-12} using the method of contradiction. Therefore, we hope to find a subset of the integral region in \eqref{III-25} that satisfies the followings: i) contains the nearest point to $\boldsymbol{\mu}_X$; ii) can be arbitrarily small so that we can use Taylor series to simplify the Gaussian integrand; iii) results in closed-form expression of the approximate integral.

According to the theorem in \eqref{III-12}, eq. \eqref{III-25} has an asymptotic expression as
\begin{equation}\label{III-26}
  \begin{split}
{P_{out}^{SC}}\left( {{\gamma _{th}}} \right) &= \int\limits_{a{x_l} + \sum\limits_{k \ne l} {{x_k}}  < \ln {\sqrt{\gamma _{th}}},\forall l=1,\cdots,L,|\textbf{x}-\boldsymbol{\mu}_X|<|\textbf{x}_0-\boldsymbol{\mu}_X|} {\frac{1}{{\sqrt {{{\left( {2\pi } \right)}^L}\sigma _X^{2L}} }}\exp \left( { - \frac{1}{{2\sigma _X^2}}|\textbf{x}-\boldsymbol{\mu}_X|^2 } \right)} d{\textbf{x}}\\&+o\left(\cdots\right)\\
%&\approx\int\limits_{a{x_l} + \sum\limits_{k \ne l} {{x_k}}  < \ln {\gamma _{th}},\forall l=1,\cdots,L,|\textbf{x}-\boldsymbol{\mu}_X|<|\textbf{x}_0-\boldsymbol{\mu}_X|} {\frac{1}{{\sqrt {{{\left( {2\pi } \right)}^L}\sigma _X^{2L}} }}\exp \left( { - \frac{1}{{2\sigma _X^2}}|\textbf{x}-\boldsymbol{\mu}_X|^2 } \right)} d{\textbf{x}}
\end{split}
\end{equation}
as long as the dominant term has a nonempty integral region where $o(\cdots)$ denotes an infinitely small quantities compared to the other summed terms when $\mu_X \to \infty$. It is proved in Appendix \ref{KKT} that the nearest point to $\boldsymbol{\mu}_X$ inside the integral region in \eqref{III-25} is $\textbf{x}_{nst}^{SC}=[\frac{{\ln \sqrt {{\gamma _{th}}} }}{{a + L - 1}},\cdots,\frac{{\ln \sqrt {{\gamma _{th}}} }}{{a + L - 1}}]^T$ and $x_{nst}^{SC} \buildrel \Delta \over = \frac{{\ln \sqrt {{\gamma _{th}}} }}{{a + L - 1}}$. We let the $L\times 1$ vector $\textbf{x}_0=[{\frac{{\ln \sqrt {{\gamma _{th}}} }}{{a + L - 1}} - \varepsilon },\cdots,{\frac{{\ln \sqrt {{\gamma _{th}}} }}{{a + L - 1}} - \varepsilon }]^T$ in \eqref{III-26} where $\varepsilon >0$, which ensures that the integral region of the dominant term in \eqref{III-26} contains $\textbf{x}_0$ and has nonzero volume, and we comment that the volume can be set arbitrarily small by adjusting $\varepsilon$, which is shown in Fig. \ref{Small_region}. It is proved in Appendix \ref{Subsets} that the integral region in \eqref{III-26} is contained by another region, i.e.
\begin{equation}\label{III-27}
\begin{split}
  \left\{\textbf{x}\left|{a{x_l} + \sum\limits_{k=1,k \ne l}^L {{x_k}}  < \ln {\sqrt{\gamma _{th}}},\forall l=1,\cdots,L,|\textbf{x}-\boldsymbol{\mu}_X|<|\textbf{x}_0-\boldsymbol{\mu}_X|}\right.\right\}\\
  \subset\left\{\textbf{x}\left| \ln {\sqrt{\gamma _{th}}} - L\left( {a + L - 1} \right)\varepsilon<a{x_l} + \sum\limits_{k=1,k \ne l}^L {{x_k}}  < \ln {\sqrt{\gamma _{th}}} ,\forall l = 1, \cdots ,L\right.\right\}
  \end{split}
\end{equation}
when $\mu_X\to \infty$. Therefore, by replacing the integral region of the dominant term in \eqref{III-26} with the larger region in \eqref{III-27}, we obtain
\begin{equation}\label{III-28}
  \begin{split}
{P_{out}^{SC}}\left( {{\gamma _{th}}} \right) &= \int\limits_{ \ln{\sqrt{\gamma _{th}}} - \delta< a{x_l} + \sum\limits_{k \ne l} {{x_k}}  <  \ln{\sqrt{\gamma _{th}}}  ,\forall l = 1, \cdots ,L} {\frac{1}{{\sqrt {{{\left( {2\pi } \right)}^L}\sigma _X^{2L}} }}\exp \left( { - \frac{1}{{2\sigma _X^2}}|\textbf{x}-\boldsymbol{\mu}_X|^2 } \right)} d{\textbf{x}}+o\left(\cdots\right)
\end{split}
\end{equation}
where $\delta\buildrel \Delta \over =L\left( {a + L - 1} \right)\varepsilon$ can be arbitrarily small when $\varepsilon$ is sufficiently small. Note that the integral regions in \eqref{III-26} and \eqref{III-28} are both subsets of the integral region in \eqref{III-25}. Figure \ref{Small_region} illustrates the relationship of the two regions in \eqref{III-27} on a two-dimensional plane.
\begin{figure}
  \centering
  \includegraphics[width=0.4\textwidth]{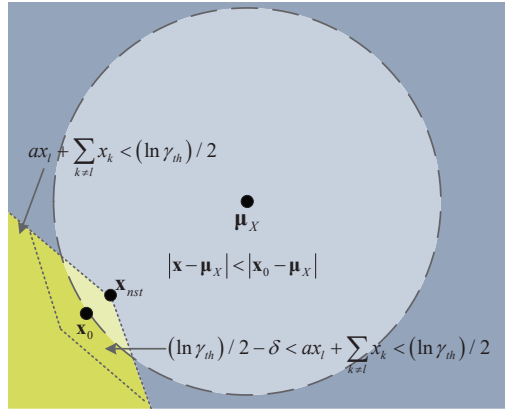}\\
  \caption{A diagram illustrating the relationship of the regions in \eqref{III-27}.}\label{Small_region}
\end{figure}

In any small neighbourhood of $\textbf{x}_{nst}^{SC}$, the integrand in \eqref{III-28} can be well approximated as
\begin{equation}\label{III-29}
\begin{split}
  &{\frac{1}{{\sqrt {{{\left( {2\pi } \right)}^L}\sigma _X^{2L}} }}\exp \left( { - \frac{1}{{2\sigma _X^2}}|\textbf{x}-\boldsymbol{\mu}_X|^2 } \right)}\\=&{\frac{1}{{\sqrt {{{\left( {2\pi } \right)}^L}\sigma _X^{2L}} }}\exp \left( { - \frac{1}{{2\sigma _X^2}}\sum\limits_{l = 1}^L {{{\left( {\left( {{x_l} - {x_{nst}^{SC}}} \right) + \left( {{x_{nst}^{SC}} - {\mu _X}} \right)} \right)}^2}} } \right)}\\
=&{\frac{1}{{\sqrt {{{\left( {2\pi } \right)}^L}\sigma _X^{2L}} }}\exp \left( { - \frac{1}{{2\sigma _X^2}}\sum\limits_{l = 1}^L {\left(2\left( {{x_l} - {x_{nst}^{SC}}} \right)\left( {{x_{nst}^{SC}} - {\mu _X}} \right){\rm{ + }}{{\left( {{x_{nst}^{SC}} - {\mu _X}} \right)}^2}+o(\cdots)\right)} } \right)}
\end{split}
\end{equation}
when $|x_l-x_{nst}^{SC}|\to 0,\forall l=1,\cdots,L$. Substituting \eqref{III-29} into \eqref{III-28}, we obtain
\begin{equation}\label{III-30}
\begin{split}
  &{P_{out}^{SC}}\left( {{\gamma _{th}}} \right){\rm{ = }}\frac{1}{{\sqrt {{{\left( {2\pi } \right)}^L}\sigma _X^{2L}} }}\exp \left( { - \frac{L}{{2\sigma _X^2}}{{\left( {{x_{nst}^{SC}} - {\mu _X}} \right)}^2}} \right)\\ &\times\int\limits_{\ln{\sqrt{\gamma _{th}}} - \delta  \le a{x_l} + \sum\limits_{k \ne l} {{x_k}}  \le \ln{\sqrt{\gamma _{th}}},\forall l=1,\cdots,L} {\exp \left( { - \frac{{{x_{nst}^{SC}} - {\mu _X}}}{{\sigma _X^2}}\sum\limits_{l = 1}^L {\left( {{x_l} - {x_{nst}^{SC}}} \right)}  + o\left(  \cdots  \right)} \right)} d{\textbf{x}}
  \end{split}
\end{equation}
when $\delta\to 0$. Then we change the integrating variables $x_l$'s to $g_l$'s in \eqref{III-30} following the mapping rule
\begin{equation}\label{III-31}
{\textbf{g}} = {\textbf{Ax}}
\end{equation}
where ${\textbf{g}}=[g_1,\cdots,g_L]^T$, ${\textbf{x}}=[x_1,\cdots,x_L]^T$ and
\begin{equation}\label{III-32}
  {\textbf{A}} = \left[ {\begin{array}{*{20}{c}}
a&1&1&1\\
1&a&1&1\\
1&1& \ddots & \vdots \\
1&1& \cdots &a
\end{array}} \right].
\end{equation}
Based on \eqref{III-31}, the integral in \eqref{III-30} can be simplified as
\begin{equation}\label{III-33}
\begin{split}
  &\int\limits_{ \ln{\sqrt{\gamma _{th}}}  - \delta  \le a{x_l} + \sum\limits_{k \ne l} {{x_k}}  \le  \ln{\sqrt{\gamma _{th}}} ,\forall l=1,\cdots,L} {\exp \left( { - \frac{{{x_{nst}^{SC}} - {\mu _X}}}{{\sigma _X^2}}\sum\limits_{l = 1}^L {\left( {{x_l} - {x_{nst}^{SC}}} \right)}  + o\left(  \cdots  \right)} \right)} d{\textbf{x}}\\
  \approx&\int\limits_{ \ln{\sqrt{\gamma _{th}}}  - \delta  \le {g_l} \le  \ln{\sqrt{\gamma _{th}}} ,\forall l=1,\cdots,L} {\exp \left( { - \frac{{{x_{nst}^{SC}} - {\mu _X}}}{{\sigma _X^2({a + L - 1})}}\sum\limits_{l = 1}^L {{g_l}}  + \frac{{{x_{nst}^{SC}} - {\mu _X}}}{{\sigma _X^2}}L{x_{nst}^{SC}}} \right)} \left| {\frac{{\partial {\textbf{x}}}}{{\partial {\textbf{g}}}}} \right|d{\textbf{g}}\\
  =&\exp \left( {\frac{{{x_{nst}^{SC}} - {\mu _X}}}{{\sigma _X^2}}L{x_{nst}^{SC}}} \right){\left| {\textbf{A}} \right|^{ - 1}}\int\limits_{ \ln{\sqrt{\gamma _{th}}}  - \delta  \le {g_l} \le  \ln{\sqrt{\gamma _{th}}} ,\forall l \le L} {\exp \left( { - \frac{{{x_{nst}^{SC}} - {\mu _X}}}{{\sigma _X^2\left( {a + L - 1} \right)}}\sum\limits_{l = 1}^L {{g_l}} } \right)} d{\textbf{g}}
  \end{split}
\end{equation}
when $\delta\to 0$, where $\left| {\frac{{\partial {\textbf{x}}}}{{\partial {\textbf{g}}}}} \right|=\left|\textbf{A}\right|^{-1}$ is the Jacobian determinant. Noting that the integrand and integral region are symmetrical for all $g_l$'s, the integral in \eqref{III-33} can be further simplified to
\begin{equation}\label{III-34}
\begin{split}
&\int\limits_{\ln{\sqrt{\gamma _{th}}} - \delta  \le {g_l} \le \ln{\sqrt{\gamma _{th}}},\forall l \le L} {\exp \left( { - \frac{{{x_{nst}^{SC}} - {\mu _X}}}{{\sigma _X^2\left( {a + L - 1} \right)}}\sum\limits_{l = 1}^L {{g_l}} } \right)} d{\textbf{g}}\\
  =&{\left[ {\int\limits_{\ln{\sqrt{\gamma _{th}}} - \delta  }^{\ln{\sqrt{\gamma _{th}}}} {\exp \left( { - \frac{{{x_{nst}^{SC}} - {\mu _X}}}{{\sigma _X^2\left( {a + L - 1} \right)}}{g_1}}\right)} d{g_1}} \right]^L}={\left( -{\frac{{\sigma _X^2\left( {a + L - 1} \right)}}{{{x_{nst}^{SC}} - {\mu _X}}}} \right)^L}\\ &\times{\left[ {\exp \left( { - \frac{{{x_{nst}^{SC}} - {\mu _X}}}{{2\sigma _X^2\left( {a + L - 1} \right)}}\ln {\gamma _{th}}} \right) - \exp \left( { - \frac{{{x_{nst}^{SC}} - {\mu _X}}}{{\sigma _X^2\left( {a + L - 1} \right)}}\left( {\frac{\ln {\gamma _{th}}}{2} - \delta } \right)} \right) } \right]^L}\\
  \approx&{\left( { - \frac{{\sigma _X^2\left( {a + L - 1} \right)}}{{{x_{nst}^{SC}} - {\mu _X}}}\exp \left( { - \frac{{{x_{nst}^{SC}} - {\mu _X}}}{{2\sigma _X^2\left( {a + L - 1} \right)}}\ln {\gamma _{th}}} \right)} \right)^L}
  \end{split}
\end{equation}
when $\mu_X \to \infty$, where the last approximation is obtained by discarding the higher order term involving $\delta$.
Substituting \eqref{III-34} into \eqref{III-33}, we can simplify \eqref{III-30} and obtain
\begin{equation}\label{III-35}
\begin{split}
  &{P_{out}^{SC}}\left( {{\gamma _{th}}} \right)\approx\frac{1}{{\left| {\bf{A}} \right|\sqrt {{{\left( {2\pi } \right)}^L}\sigma _X^{2L}} }}{\left( {\frac{{\sigma _X^2\left( {a + L - 1} \right)}}{{{\mu _X} - \frac{{\ln {\gamma _{th}}}}{{2(a+L - 1)}}}}} \right)^L}\exp \left( { - \frac{L}{{2\sigma _X^2}}{{\left( {\frac{{\ln {\gamma _{th}}}}{{2(a{\rm{ + }}L - 1)}} - {\mu _X}} \right)}^2}} \right)
  \end{split}
\end{equation}
when $\mu_X\to \infty$. Noting that ${\mu _X} = \frac{{\ln \sqrt {{{\bar E}_r}}  - \sigma _G^2}}{{a + L - 1}}$ and $\sigma _X^2 = \frac{{\sigma _G^2}}{{{a^2} + L - 1}}$, we can also express \eqref{III-35} using the standard deviation $\sigma_G$ and the transmit power ${{\bar E}_r}$ as\footnote{We keep some higher order small quantities to make the asymptotic expression converge faster.}
\begin{equation}\label{III-36}
\begin{split}
  {P_{out}^{SC}}\left( {{\gamma _{th}}} \right)&\approx
  \frac{1}{{\left| {\bf{A}} \right|\sqrt {{{\left( {2\pi } \right)}^L}} }}{\left( {\frac{{\sigma _G^2}}{{{a^2} + L - 1}}} \right)^{\frac{L}{2}}}{\left( {\frac{{{{\left( {a + L - 1} \right)}^2}}}{{\ln \sqrt {{{\bar E}_r}/{\gamma _{th}}}  - \sigma _G^2}}} \right)^L}\\ &\times\exp \left( { - \frac{{L\left( {{a^2} + L - 1} \right)}}{{2\sigma _G^2}}{{\left( {\frac{{\ln \sqrt {{{\bar E}_r}/{\gamma _{th}}}  - \sigma _G^2}}{{a + L - 1}}} \right)}^2}} \right)
  \end{split}
\end{equation}
when ${{\bar E}_r}\to \infty$.

When $a\to\infty$, the correlation coefficient $\rho\to 0$ according to \eqref{II-4}, and the lognormal channels become independent. In such case, eq. \eqref{III-36} specializes to
\begin{equation}\label{III-37}
  {P_{out}^{SC,i}}\left( {{\gamma _{th}}} \right) \approx \frac{{\sigma _G^L}}{{{{\left( {\ln \sqrt {{{\bar E}_r}/{\gamma _{th}}}  - \sigma _G^2} \right)}^L}\sqrt {{{\left( {2\pi } \right)}^L}} }}\exp \left( { - \frac{L}{{2\sigma _G^2}}{{\left( {\ln \sqrt {{{\bar E}_r}/{\gamma _{th}}}  - \sigma _G^2} \right)}^2}} \right).
\end{equation}
We can also use another simpler approach to obtain \eqref{III-37} as a verification to the proposed analysis in the case of independent lognormal channels. The exact CDF of a lognormal RV $c$ is known as
\begin{equation}\label{III-38}
  F_{c}(x)=1 - Q\left( {\frac{{\ln x - \mu_{c} }}{\sigma_{c} }} \right)=Q\left( {\frac{{\mu_c  - \ln x}}{\sigma_c }} \right)
\end{equation}
where $\mu_c$ and $\sigma_c^2$ are the mean and variance of the associated Gaussian RV. Based on \eqref{II-6} and \eqref{III-38}, the exact outage probability of SC over the independent lognormal channels is
\begin{equation}\label{III-40}
  P_{out}^{SC,i}\left( {{\gamma _{th}}} \right) = {Q^L}\left( {\frac{{{\mu _G} - \ln{\sqrt{\gamma _{th}}}}}{{{\sigma _G}}}} \right) = {Q^L}\left( {\frac{{\ln \sqrt {{{\bar E}_r}/{\gamma _{th}}}  - \sigma _G^2}}{{{\sigma _G}}}} \right).
\end{equation}
Based on the well-known asymptotic approximation of Gaussian $Q$-function\cite[eq. (4)]{Qfunction_bounds}
\begin{equation}\label{III-41}
  Q\left( x \right) \approx \frac{1}{{\sqrt {2\pi } x}}\exp \left( { - \frac{{{x^2}}}{2}} \right)
\end{equation}
when $x\to\infty$, eq. \eqref{III-40} can be approximated as
\begin{equation}\label{III-42}
  P_{out}^{SC,i}\left( {{\gamma _{th}}} \right) \approx \frac{{\sigma _G^L}}{{ {{\left( {\ln \sqrt {{{\bar E}_r}/{\gamma _{th}}}  - \sigma _G^2} \right)}^L} \sqrt {(2\pi)^L }}}\exp \left( { - \frac{{L}}{{2\sigma _G^2}}}{{\left( {\ln \sqrt {{{\bar E}_r}/{\gamma _{th}}}  - \sigma _G^2} \right)}^2} \right)
\end{equation}
which agrees with \eqref{III-37}.

\section{Asymptotic Outage Probabilities of EGC and MRC Over Correlated Channels}\label{EGCMRCoutage}
Based on \eqref{II-6.2}, we can express the outage probability of EGC over equally correlated channels as
\begin{equation}\label{III-43}
  P_{out}^{EGC}\left( {{\gamma _{th}}} \right) = \int\limits_{ \sum\limits_{l = 1}^L {\exp \left( {a{x_l} + \sum\limits_{k=1,k \ne l}^L {{x_k}} } \right)}  \le \sqrt {L{\gamma _{th}}}} {{f_{{\textbf{x}},iid}}\left( {\textbf{x}} \right)d{\textbf{x}}}
\end{equation}
where we assume $\boldsymbol{\mu}=\boldsymbol{\mu}_X$. The integral region in \eqref{III-43} can be expressed as
\begin{equation}\label{III-44}
  \Phi_{EGC} \left( {\textbf{x}} \right) \buildrel \Delta \over = \sum\limits_{l = 1}^L {\exp \left( {a{x_l} + \sum\limits_{k=1,k \ne l}^L {{x_k}} } \right)} - \sqrt {L{\gamma _{th}}}\le0.
\end{equation}
According to the theorem in \eqref{III-12}, we obtain
\begin{equation}\label{III-45}
  \int\limits_{\Phi_{EGC} \left( {\textbf{x}} \right) \le 0} {{f_{{\textbf{x}},iid}}\left( {\textbf{x}} \right)d{\textbf{x}}}  = \int\limits_{\Phi_{EGC} \left( {\textbf{x}} \right) \le 0,\left| {{\textbf{x}} - {\boldsymbol{\mu }}_X} \right| < \left| {{{\bf{x}}_{nst}^{EGC}} - {\boldsymbol{\varepsilon }} - {\boldsymbol{\mu }_X}} \right|} {{f_{{\textbf{x}},iid}}\left( {\textbf{x}} \right)d{\textbf{x}}}+o(\cdots)
\end{equation}
for $\mu_X\to\infty$ where $\boldsymbol{\varepsilon}=[\varepsilon,\cdots,\varepsilon]^T$ and ${{\textbf{x}}_{nst}^{EGC}}$ is the nearest point to $\boldsymbol{\mu}_X$ in the region $\Phi \left( {\textbf{x}} \right)_{EGC}\leq0$. Applying the Karush-Kuhn-Tucker (KKT) conditions, it can be shown that ${{\textbf{x}}_{nst}^{EGC}} = \left[ {\frac{1}{{a + L - 1}}\ln \left( {\sqrt {\frac{{{\gamma _{th}}}}{L}} } \right), \cdots ,\frac{1}{{a + L - 1}}\ln \left( {\sqrt {\frac{{{\gamma _{th}}}}{L}} } \right)} \right]^T$, and the procedures are similar to those in Appendix \ref{KKT}. The integral region $\Phi_{EGC} \left( {\textbf{x}} \right)\leq0$ in \eqref{III-44} can be approximated by the following hyperspherical region
\begin{equation}\label{III-46}
  \tilde \Phi_{EGC} \left( {\textbf{x}} \right) \buildrel \Delta \over ={\sum\limits_{l = 1}^L {{{\left( {{x_l} - \left( {\frac{1}{{a + L - 1}}\ln \left( {\sqrt {\frac{{{\gamma _{th}}}}{L}} } \right) - \frac{{\left( {L - 1 + a} \right)}}{{{{\left( {1 - a} \right)}^2}}}} \right)} \right)}^2}} - {{\left( {\frac{{\left( {L - 1 + a} \right)\sqrt L }}{{{{\left( {1 - a} \right)}^2}}}} \right)}^2}}\le0
\end{equation}
where the approximation is valid because if we take $x_1$ as a function of $x_2,\cdots,x_L$, for hypersurface $\Phi_{EGC} \left( {\textbf{x}} \right)=0$ and $\tilde \Phi_{EGC} \left( {\textbf{x}} \right)=0$, the first-order partial derivatives $\frac{{\partial {x_1}}}{{\partial {x_m}}},\forall m=1,\cdots,L$ at ${{\textbf{x}}_{nst}^{EGC}}$ are identical, and so do the second-order partial derivatives $\frac{{{\partial ^2}{x_1}}}{{\partial {x_m}\partial {x_n}}},\forall m,n =2,\cdots,L$. The proof of the equalities of derivatives is in Appendix \ref{Derivatives}. This implies that the two hypersurfaces $\Phi_{EGC} \left( {\textbf{x}} \right)=0$ and $\tilde \Phi_{EGC} \left( {\textbf{x}} \right)=0$ has arbitrarily small difference in a sufficiently small neighbourhood of $\textbf{x}_{nst}^{EGC}$. One can also regard the approximation as a generalization of the curve approximation technique using a circle that has the same curvature. For $L=2$, the reciprocal of the radius of \eqref{III-46}, i.e. $\frac{{{{\left( {1 - a} \right)}^2}}}{{\left( {L - 1 + a} \right)\sqrt L }}$, is the curvature of the curve at $\textbf{x}_{nst}^{EGC}$.

Therefore, according to the theorem in \eqref{III-12}, eq. \eqref{III-45} can be approximated as
\begin{equation}\label{III-47}
\begin{split}
\int\limits_{\Phi_{EGC} \left( {\textbf{x}} \right) \le 0} {{f_{{\textbf{x}},iid}}\left( {\textbf{x}} \right)d{\textbf{x}}}  &\approx \int\limits_{ \tilde \Phi_{EGC} \left( {\textbf{x}} \right) \le 0,\left| {{\textbf{x}} - {\boldsymbol{\mu }}_X} \right| < \left| {{{\textbf{x}}_{nst}^{EGC}} - {\boldsymbol{\varepsilon }} - {\boldsymbol{\mu }_X}} \right|} {{f_{{\textbf{x}},iid}}\left( {\textbf{x}} \right)d{\textbf{x}}}\approx\int\limits_{ \tilde \Phi_{EGC} \left( {\textbf{x}} \right) \le 0} {{f_{{\textbf{x}},iid}}\left( {\textbf{x}} \right)d{\textbf{x}}}.
 \end{split}
\end{equation}
when $\mu_X \to \infty$. An intuitive example for the approximation in \eqref{III-47} is shown Fig. \ref{subfigs}, where the integral region is $\Phi_{EGC}(\textbf{x})\leq0$ for $L=2$. Since the dominant term of the integrand is condensed into a small region, we only need to accurately approximate the integral boundary near the dominant term. Based on \eqref{III-46}, the last integral in \eqref{III-47} can be expressed as
\begin{equation}\label{III-48}
\begin{split}
  &P_{out}^{EGC}\left( {{\gamma _{th}}} \right)\approx\int\limits_{ \tilde \Phi_{EGC} \left( {\textbf{x}} \right) \le 0} {{f_{{\textbf{x}},iid}}\left( {\textbf{x}} \right)d{\textbf{x}}}\\
  &=\Pr \left\{ {\sum\limits_{l = 1}^L {{{\left( {{X_l} - \left( {\frac{1}{{a + L - 1}}\ln \left( {\sqrt {\frac{{{\gamma _{th}}}}{L}} } \right) - \frac{{\left( {L - 1 + a} \right)}}{{{{\left( {1 - a} \right)}^2}}}} \right)} \right)}^2}}  \le {{\left( {\frac{{\left( {L - 1 + a} \right)\sqrt L }}{{{{\left( {1 - a} \right)}^2}}}} \right)}^2}} \right\}\\
  & = \Pr \left\{ {\sum\limits_{l = 1}^L {{\left( \underbrace{{\frac{{{X_l} - \left( {\frac{1}{{a + L - 1}}\ln \left( {\sqrt {\frac{{{\gamma _{th}}}}{L}} } \right) - \frac{{\left( {L - 1 + a} \right)}}{{{{\left( {1 - a} \right)}^2}}}} \right)}}{{{\sigma _X}}}}}_{Y_l} \right)}}^2  \le {{\left( {\frac{{\left( {L - 1 + a} \right)\sqrt L }}{{{{\left( {1 - a} \right)}^2}{\sigma _X}}}} \right)}^2}} \right\}
  \end{split}
\end{equation}
where $Y_l\sim {\cal{N}} \left({\frac{{{\mu_X} - \left( {\frac{1}{{a + L - 1}}\ln \left( {\sqrt {\frac{{{\gamma _{th}}}}{L}} } \right) - \frac{{\left( {L - 1 + a} \right)}}{{{{\left( {1 - a} \right)}^2}}}} \right)}}{{{\sigma _X}}}},1\right)$. $\sum\limits_{l = 1}^L {Y_l^2} $ follows the $L$th-order noncentral chi-squared distribution whose CDF can be expressed as Marcum-$Q$ function defined as ${Q_M}\left( {a,b} \right) = \int\limits_b^\infty  x{{\left( {\frac{x}{a}} \right)}^{M - 1}}$ $\exp \left( { - \frac{{{x^2} + {a^2}}}{2}} \right){I_{M - 1}}\left( {ax} \right)dx $ where $I_{M - 1}(x)$ is the modified Bessel function of order $M-1$, thus we can simplify \eqref{III-48} as
\begin{equation}\label{III-48.5}
\begin{split}
&P_{out}^{EGC}\left( {{\gamma _{th}}} \right)\approx1 - {Q_{\frac{L}{2}}}\left( {\sqrt L \left( {\frac{{{\mu _X} - \left( {\frac{1}{{a + L - 1}}\ln \left( {\sqrt {\frac{{{\gamma _{th}}}}{L}} } \right) - \frac{{L - 1 + a}}{{{{\left( {1 - a} \right)}^2}}}} \right)}}{{{\sigma _X}}}} \right),\frac{{\left( {L - 1 + a} \right)\sqrt L }}{{{{\left( {1 - a} \right)}^2}{\sigma _X}}}} \right).
\end{split}
\end{equation}
Equation \eqref{III-48.5} can also be expressed as a function of the average received power ${{{\bar E}_r}}$ and the standard deviation $\sigma_G$ as
\begin{equation}\label{III-49}
\begin{split}
&P_{out}^{EGC}(\gamma_{th})\approx  1 - {Q_{\frac{L}{2}}}\left( {\sqrt L \left( {\frac{{\frac{{\ln \sqrt {{{L\bar E}_r}/{\gamma _{th}}}  - \sigma _G^2  }}{{a + L - 1}} + \frac{{L - 1 + a}}{{{{\left( {1 - a} \right)}^2}}}}}{{\frac{{{\sigma _G}}}{{\sqrt {{a^2} + L - 1} }}}}} \right),\frac{{\left( {L - 1 + a} \right)\sqrt L }}{{{{\left( {1 - a} \right)}^2}\frac{{{\sigma _G}}}{{\sqrt {{a^2} + L - 1} }}}}} \right).
\end{split}
\end{equation}
When $a\to\infty$, we obtain $\rho\to 0$ according to \eqref{II-4}, thus the channels become independent and \eqref{III-49} becomes
\begin{equation}\label{III-49.5}
  P_{out}^{EGC,i}\left( {{\gamma _{th}}} \right) \approx 1 - {Q_{\frac{L}{2}}}\left( {\frac{{\sqrt L }}{{{\sigma _G}}}\left( {\ln \sqrt {L{{\bar E}_r}/{\gamma _{th}}}  - \sigma _G^2 + 1} \right),\frac{{\sqrt L }}{{{\sigma _G}}}} \right)
\end{equation}
which agrees with the prior result in \cite[eq. (14)]{IndependentLognormalMRCEGC}.

According to \eqref{II-6.4}, the associated integral region for MRC can be expressed as
\begin{equation}\label{III-49.8}
   \Phi_{MRC} \left( {\textbf{x}} \right) \buildrel \Delta \over = \sum\limits_{l = 1}^L {\exp \left( {2\left( {a{x_l} + \sum\limits_{k \ne l} {{x_k}} } \right)} \right)} - {\gamma _{th}}\le0
\end{equation}
and using the same technique in Appendix \ref{Derivatives}, the integral region to approximate $\Phi_{MRC} \left( {\textbf{x}} \right)$ can be derived as
\begin{equation}\label{III-49.9}
  \tilde \Phi_{MRC} \left( {\textbf{x}} \right) \buildrel \Delta \over ={\sum\limits_{l = 1}^L {{{\left( {{x_l} - \left( {\frac{1}{{a + L - 1}}\ln \left( { {\frac{{{\gamma _{th}}}}{L}} } \right) - \frac{{\left( {L - 1 + a} \right)}}{{{{\left( {1 - a} \right)}^2}}}} \right)} \right)}^2}} - {{\left( {\frac{{\left( {L - 1 + a} \right)\sqrt L }}{{{{\left( {1 - a} \right)}^2}}}} \right)}^2}}\le0.
\end{equation}
By comparing \eqref{II-6.2} and \eqref{II-6.4}, it can be observed that the outage probability of MRC is also in form of the CDF of sum of lognormal RVs like that of EGC.
Therefore, noting that $2X_l\sim{\cal{N}}(2\mu_X,4\sigma_X^2),\forall l=1,\cdots,L$, we can obtain
\begin{equation}\label{III-50}
  P_{out}^{MRC}\left( {\gamma _{th}'} \right) \approx 1 - {Q_{\frac{L}{2}}}\left( {\sqrt L \left( {\frac{{2\mu _X' - \left( {\frac{1}{{a + L - 1}}\ln \left( {\frac{{\gamma _{th}'}}{L}} \right) - \frac{{L - 1 + a}}{{{{\left( {1 - a} \right)}^2}}}} \right)}}{{2\sigma _X'}}} \right),\frac{{\left( {L - 1 + a} \right)\sqrt L }}{{{{\left( {1 - a} \right)}^2}2\sigma _X'}}} \right)
\end{equation}
by letting $\sigma_X^2=4\sigma_X'^2$, $\mu_X=2\mu_X'$ and $\gamma_{th}'=\sqrt{L\gamma_{th}}$ in \eqref{III-48.5}. Equation \eqref{III-50} can also be expressed as a function of $\bar E_r$ as
\begin{equation}\label{III-51}
   P_{out}^{MRC}\left( {{\gamma _{th}}} \right) \approx  1 - {Q_{\frac{L}{2}}}\left( {\sqrt L \left( {\frac{{\frac{{\ln \left( {{{L\bar E}_r}/{\gamma _{th}}} \right) - 2\sigma _G^2 }}{{a + L - 1}} + \frac{{L - 1 + a}}{{{{\left( {1 - a} \right)}^2}}}}}{{2\frac{{{\sigma _G}}}{{\sqrt {{a^2} + L - 1} }}}}} \right),\frac{{\left( {L - 1 + a} \right)\sqrt L }}{{{{\left( {1 - a} \right)}^2}\frac{{2{\sigma _G}}}{{\sqrt {{a^2} + L - 1} }}}}} \right).
\end{equation}
Taking $a\to\infty$, we obtain the MRC outage probability for independent channels as
\begin{equation}\label{III-52}
  P_{out}^{MRC,i}\left( {{\gamma _{th}}} \right) \approx 1 - {Q_{\frac{L}{2}}}\left( {\frac{{\sqrt L }}{{2{\sigma _G}}}\left( {\ln \left( {L{{\bar E}_r}/{\gamma _{th}}} \right) - 2\sigma _G^2 + 1} \right),\frac{{\sqrt L }}{{2{\sigma _G}}}} \right)
\end{equation}
which agrees with the prior result in \cite[eq. (16)]{IndependentLognormalMRCEGC}.

\section{Discussion On The Asymptotic Outage Probabilities}\label{discussions}
\subsection{Comparison Between Lognormal Fading Channels and Other Channels}\label{compareChannel}
For most fading channels, such as Rayleigh, Rician, Nakagami-$m$, Weibull, $\alpha$-$\mu$, Gamma-Gamma, Negative-exponential etc., the asymptotic outage probability can be expressed as \cite{wang2003simple}
\begin{equation}\label{IV-1}
  {P_{out}}\left( {{\gamma _{th}}} \right) \approx {\left( {{O_c}{{\bar E}_r}} \right)^{ - {O_d}}}
\end{equation}
where $O_c$ and $O_d$ are, respectively, known as the coding gain and the diversity order. Taking logarithm operation on both sides of \eqref{IV-1}, we obtain
\begin{equation}\label{IV-2}
  \lg {P_{out}}\left( {{\gamma _{th}}} \right){\rm{ = }} - {O_d}\lg \left( {{O_c}} \right) - \frac{{{O_d}}}{{10}}\left( {10\lg  {{{\bar E}_r}}} \right)
\end{equation}
where ${10\lg  {{{\bar E}_r}} }$ is the average received power in dB. Equation \eqref{IV-2} illustrates that on logarithmic coordinates the outage probability has a straight asymptote whose slope is determined by $O_d$ and the horizontal shift is determined by $O_c$. For a diversity reception system, prior works \cite{BeckmannBingcheng,BingchengHierarchical,XueguiDiversity} proved that channel correlation and the combining schemes (MRC, SC and EGC) only influence $O_c$ for the channels with finite diversity orders.

Taking the same logarithm operation on \eqref{III-36}, we can obtain
\begin{equation}\label{IV-3}
  \lg {P_{out}^{SC}}\left( {{\gamma _{th}}} \right) = \lg {O_{c,ln}^{SC}} - L\lg \left(\ln  {{{\sqrt{\bar E_r/\gamma_{th}}}}} -\sigma_G^2\right) - {O_{d,ln}^{SC}}{\left( {\ln {\sqrt {\bar E_r /{\gamma _{th}}}}  - \sigma _G^2} \right)^2}
\end{equation}
where $O_{c,ln}^{SC} \buildrel \Delta \over = \frac{{{{\left( {a + L - 1} \right)}^{2L}}}}{{\left| {\textbf{A}} \right|\sqrt {{{\left( {2\pi } \right)}^L}} }}{\left( {\frac{{\sigma _G^2}}{{{a^2} + L - 1}}} \right)^{\frac{L}{2}}}$ and $O_{d,ln}^{SC}\buildrel \Delta \over ={  \frac{{\lg (e) L\left( {{a^2} + L - 1} \right)}}{{2\sigma _G^2{{\left( {a + L - 1} \right)}^2}}}}$.

It can be observed from \eqref{IV-3} that the asymptote of the outage probability of SC is not a straight line on logarithmic coordinates. Similar to that in \eqref{IV-2}, the first term $O_{c,ln}^{SC}$ determines the shift of the asymptote, which is related to the correlation coefficient $\rho$, branch number $L$ and the variance $\sigma_G^2$. The second term in \eqref{IV-3} is related to $L$, $\sigma_G^2$, $\gamma_{th}$ and $\bar E_r$, and the double-fold logarithm operation $\lg\ln(\cdot)$ greatly suppresses the increase of $\bar E_r$, thus the second term will not induce significant drop in outage probability as $\bar E_r$ increases. For example, $\lg\ln(10^5)=1.06$ and $\lg\ln(10^{15})=1.53$. In contrast, the third term in \eqref{IV-3} is a second-order polynomial of $\ln\bar E_r$, which will introduce dramatic decrease in the outage probability when $\bar E_r$ increases. For example, $\ln(10^1)^2=5.3019$ and $\ln(10^2)^2=21.2076$. The scale $O_{d,ln}^{SC}$ of the second-order polynomial will dominantly determine the dropping speed of the outage probability. If an SC system has higher $O_{d,ln}^{SC}$ than the other SC systems, it will inevitably have lower outage probability when the signal power is sufficiently large. More importantly, since $O_{d,ln}^{SC}$ is a function of $a$, or the correlation coefficient $\rho$, the decreasing speed of the asymptote of the outage probability is related to $\rho$. In contrast, the slope of the asymptote is fixed for the other channels with finite diversity order for different correlation status.

For EGC and MRC, the asymptotic outage probabilities are expressed using Marcum-$Q$ function in \eqref{III-49} and \eqref{III-51}, and it is not straightforward to perform a similar expansion as we did in \eqref{IV-3}. However, in \cite{Concavity_MarcumQ}, the authors proved the property of log-concavity of function $1-Q_v(p,q)$ with respect to $p$, which leads to the conclusion that $\lg P_{out}^{EGC}(\gamma_{th})$ and $\lg P_{out}^{MRC}(\gamma_{th})$ are both log-concave with respect to $\ln \bar E_r$, or $10 \lg \bar E_r$ (the average branch SNR in dB) according to \eqref{III-49} and \eqref{III-51}. This result predicts that there is no straight asymptote for the outage probabilities of MRC and EGC over lognormal fading channels on logarithmic coordinates, and the  outage probabilities drop dramatically fast in high SNR region. Besides, eqs. \eqref{III-49} and \eqref{III-51} both show that $\ln \bar E_r$ is scaled by a factor related to $a$, $L$ and $\sigma_X$, which illustrates that the dropping speed of the outage probabilities is related to the correlation coefficient $\rho$, branch number $L$ and $\sigma_G$. In contrast, the diversity orders of MRC, EGC and SC over Rayleigh, Rician and Nakagami-$m$ channels are not related to the channel correlation in large SNR region \cite{XueguiDiversity,BingchengNakagami,wang2003simple}.

\subsection{Comparison Between SC, MRC And EGC}\label{SCcomparison}

\begin{figure}
  \centering
  \includegraphics[width=0.6\textwidth]{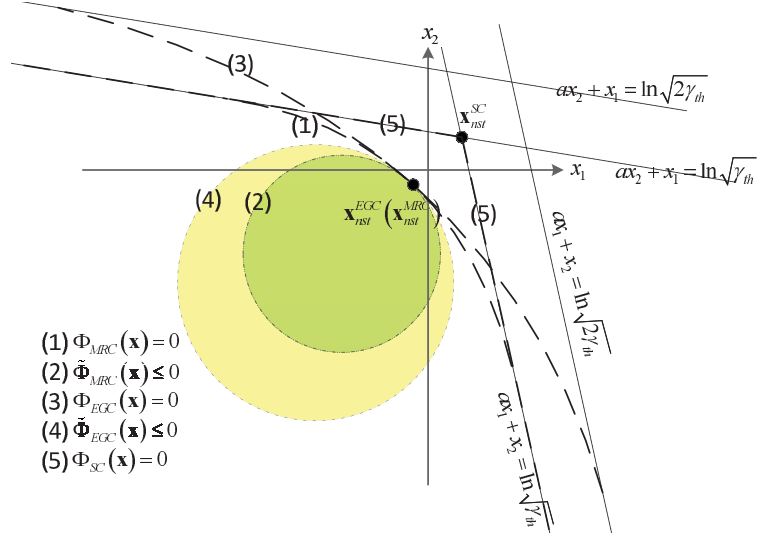}\\
  \caption{Integral regions $\tilde \Phi_{EGC}(\textbf{x})\leq0$, $\tilde \Phi_{MRC}(\textbf{x})\leq0$, $\Phi_{SC}(\textbf{x})\leq0$, $\Phi_{EGC}(\textbf{x})\leq0$ and $\Phi_{MRC}(\textbf{x})\leq0$ on a two-dimensional plane. The asymptotes for the boundary $\Phi_{MRC}(\textbf{x})=0$ are $  {a{x_l} + \sum\limits_{k \ne l} {{x_k}} }  = \ln \sqrt {{\gamma _{th}}}, \forall l=1,\cdots,L$. The asymptotes for the boundary $\Phi_{EGC}(\textbf{x})=0$ are $ {a{x_l} + \sum\limits_{k \ne l} {{x_k}} }  = \ln \sqrt {{L\gamma _{th}}}, \forall l=1,\cdots,L$. The nearest points for regions $\Phi_{EGC}(\textbf{x})\leq0$ and $\Phi_{MRC}(\textbf{x})\leq0$ are identical.}\label{DiversityRegions}
\end{figure}

The theorem in \eqref{III-12} reveals that the asymptotic outage probability is determined by the nearest point to $\boldsymbol{\mu}_X$ in the integral region. The integrands of the outage probabilities for SC, MRC and EGC are $f_{\textbf{x},iid}(\textbf{x})$. The integral regions of SC and EGC are shown, respectively, in \eqref{III-25.5} and \eqref{III-44}. Figure \ref{DiversityRegions} shows the relationship between the regions $\tilde \Phi_{EGC} \left( {\textbf{x}} \right)\leq0$, $\Phi_{EGC} \left( {\textbf{x}} \right)=0$, $\tilde \Phi_{MRC} \left( {\textbf{x}} \right)\leq0$, $\Phi_{MRC} \left( {\textbf{x}} \right)=0$ and $\Phi_{SC} \left( {\textbf{x}} \right)=0$ on a two-dimensional plane. In Appendix \ref{KKT}, it has been proved that the nearest point to $\boldsymbol{\mu}_X$ in $\Phi_{SC}(\textbf{x})\leq0$ is $\textbf{x}_{nst}^{SC}=\left[{\frac{{\ln \sqrt{\gamma _{th}}}}{a + L - 1} },\cdots,{\frac{{\ln \sqrt{\gamma _{th}}}}{{a + L - 1}} }\right]^T$. In $\Phi_{EGC}(\gamma_{th})\leq0$ and $\Phi_{MRC}(\gamma_{th})\leq0$, the nearest points to $\boldsymbol{\mu}_X$ can be proved to be
 \begin{equation}\label{IV-8.5}
 {{\textbf{x}}_{nst}^{EGC}} ={{\textbf{x}}_{nst}^{MRC}} = \left[ \frac{\ln  {\sqrt {{{{\gamma _{th}}}}} } -\ln \sqrt{L}}{{a + L - 1}},\cdots ,\frac{\ln  {\sqrt {{{{\gamma _{th}}}}} } -\ln \sqrt{L}}{{a + L - 1}} \right]^T
\end{equation}
which is based on the KKT conditions applied in Appendix \ref{KKT}. If we make $\textbf{x}_0 = (\textbf{x}_{nst}^{EGC}+\textbf{x}_{nst}^{SC})/2$, we can obtain
\begin{equation}\label{IV-9.4}
  \left| {{{\textbf{x}}_0} - {{\boldsymbol{\mu }}_X}} \right| = \sqrt L \left( {{\mu _X} - \frac{{\ln \sqrt {{\gamma _{th}}}  - \frac{1}{2}\ln \sqrt L }}{{a + L - 1}}} \right)
\end{equation}
when $\mu_X\to\infty$. Besides, it can also be calculated that
\begin{equation}\label{IV-9.2}
  \left| {{\textbf{x}}_{nst}^{EGC} - {{\boldsymbol{\mu }}_X}} \right| = \sqrt L \left( {{\mu _X} - \frac{{\ln \sqrt {{\gamma _{th}}}  - \ln \sqrt L }}{{a + L - 1}}} \right)
\end{equation}
and
\begin{equation}\label{IV-9.3}
  \left| {{\textbf{x}}_{nst}^{SC} - {{\boldsymbol{\mu }}_X}} \right|   = \sqrt L \left( \mu_X-\frac{{\ln \sqrt{\gamma _{th}}}}{{a + L - 1}}\right).
\end{equation}
Equations \eqref{IV-9.4}, \eqref{IV-9.2} and \eqref{IV-9.3} show
\begin{equation}\label{IV-9.5}
  \left|\textbf{x}_{nst}^{EGC}-\boldsymbol{\mu}_X\right|=\left|\textbf{x}_{nst}^{MRC}-\boldsymbol{\mu}_X\right|>\left|\textbf{x}_{0}-\boldsymbol{\mu}_X\right|>\left|\textbf{x}_{nst}^{SC}-\boldsymbol{\mu}_X\right|
\end{equation}
when $\mu_X\to\infty$. Based on \eqref{IV-9.5} and the theorem in \eqref{III-12}, we obtain
\begin{equation}\label{IV-10}
\begin{split}
  \int\limits_{{\Phi _{EGC}}\left( {{\gamma _{th}}} \right)\leq 0} {{f_{{\textbf{x}},iid}}\left( {\textbf{x}} \right)d{\textbf{x}}} & = o\left( {\int\limits_{{\Phi _{SC}}\left( {{\gamma _{th}}} \right)\leq 0,\left| {{\textbf{x}} - {{\textbf{x}}_0}} \right| < \left| {{{\boldsymbol{\mu }}_X} - {{\textbf{x}}_0}} \right|} {{f_{{\textbf{x}},iid}}\left( {\textbf{x}} \right)d{\textbf{x}}} } \right) = o\left( {\int\limits_{{\Phi_{SC}\left( {{\gamma _{th}}} \right)\leq0}} {{f_{{\textbf{x}},iid}}\left( {\textbf{x}} \right)d{\textbf{x}}} } \right)
  \end{split}
\end{equation}
when $\mu_X\to\infty$, where the first equality holds because the integral region on the right-hand side is within the region $\left| {{\textbf{x}} - {{\textbf{x}}_0}} \right| < \left| {{{\boldsymbol{\mu }}_X} - {{\textbf{x}}_0}} \right|$, and ${\Phi _{EGC}}\left( {{\gamma _{th}}} \right)\leq 0$ is outside the region $\left| {{\textbf{x}} - {{\textbf{x}}_0}} \right| < \left| {{{\boldsymbol{\mu }}_X} - {{\textbf{x}}_0}} \right|$. Similarly, we can obtain
\begin{equation}\label{IV-11}
  \int\limits_{{\Phi _{MRC}}\left( {{\gamma _{th}}} \right)\leq0} {{f_{{\textbf{x}},iid}}\left( {\textbf{x}} \right)d{\textbf{x}}}  = o\left( {\int\limits_{{\Phi _{SC}}\left( {{\gamma _{th}}} \right)\leq0} {{f_{{\textbf{x}},iid}}\left( {\textbf{x}} \right)d{\textbf{x}}} } \right)
\end{equation}
when $\mu_X\to\infty$. Equations \eqref{IV-10} and \eqref{IV-11} reveal that the performance gap between SC and EGC (or MRC) will become infinite when $\mu_X$ (or $\bar E_r$) is sufficiently large. In contrast, for the other fading channels with finite diversity orders, the performance gaps between different combining schemes are fixed in high SNR region \cite{XueguiDiversity,wang2003simple,BingchengNakagami,BeckmannBingcheng}.

\section{Numerical Results}\label{numerical}
\begin{figure}
  \centering
  \includegraphics[width=0.5\textwidth]{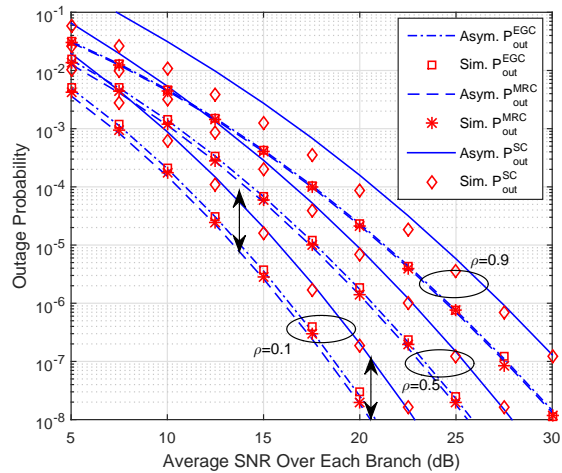}\\
  \caption{Outage probabilities of dual-branch MRC, EGC, SC over independent lognormal fading channels and the outage probability of a single-branch link.  $\sigma_G=0.8$; $\gamma_{th}=0.1$W; $\rho=0.1,0.5,0.9$. ``Asym.'' is short for ``Asymptotic'' and ``Sim.'' is short for ``Simulation''.  The two double-arrow line segments have identical length.}\label{ChangeRho}
\end{figure}

Figure \ref{ChangeRho} shows the outage probabilities of SC, EGC and MRC over lognormal channels with different correlation coefficients. It can be observed that the simulated outage probability values converge to the asymptotic outage probability for different correlation coefficients. Besides, the performance gap between strongly and weakly correlated channels increases with the transmit power, and this is in sharp contrast with the other channels (Rayleigh, Rician, Nakagami-$m$, etc.) for which the performance gap over weakly and strongly correlated channels approaches a fixed value in large SNR region. This result agrees with our analyses in \eqref{IV-3} that the dropping speed of the asymptote of outage probability at a large SNR value is determined by the value $O_{d,ln}$ which is related to the correlation coefficient $\rho$, and it also agrees with the analyses in \eqref{III-49} and \eqref{III-51} which show that $\rho$ contributes a scaler of $\ln\bar E_r$ and influences the slope at high SNR. Another conclusion that can be drawn from Fig. \ref{ChangeRho} is that the outage probabilities of SC, EGC and MRC over lognormal fading channels do not have a straight asymptote; instead, the slope of the asymptote  decreases to $-\infty$ when $\bar E_r \to \infty$. This observation verifies the prediction in Section \ref{compareChannel}. In Fig. \ref{ChangeRho}, the two double-arrow line segments have the same length, which reveals that the performance gap between SC and MRC (or EGC) increases with $\bar E_r$, and this agrees with our discussion in Section \ref{SCcomparison}. The converging speed of the asymptotic expressions for EGC and MRC is observed to be much faster than that of SC, and this is because the derivation process for SC was based on the approximation of both integral region and the integrand, while the derivation processes for EGC and MRC were based on the approximation the integral region. Besides, it can also be observed that when $\rho=0.9$, which implies that the channels are highly correlated, the performance gap between EGC and MRC is negligible. This comply with the intuition that the weights of MRC for combined branches tend to be identical for strongly correlated channels, and MRC is equivalent to EGC when $\rho=1$.

\begin{figure}
  \centering
  \includegraphics[width=0.5\textwidth]{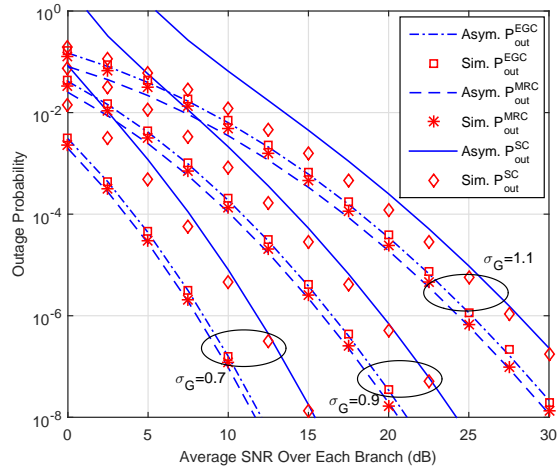}\\
  \caption{Outage probabilities of SC, EGC and MRC over correlated lognormal fading channels. The correlation coefficients are $\rho=0.2$; $\sigma_G=0.7, 0.9, 1.1$; $\gamma_{th}=0.1$W; $L=3$.}\label{ChangeSigma}
\end{figure}
Figure \ref{ChangeSigma} compares the outage probabilities of SC, EGC and MRC over lognormal fading channels with different $\sigma_G$'s. It can be observed again that the asymptotic outage probabilities well approximate the exact outage probabilities in large SNR region. Figure \ref{ChangeSigma} also reveals that the slopes of the asymptotes at a large SNR value are highly related to $\sigma_G$. For radio-frequency communications where the lognormal parameter $\sigma_G$ describes the severity of large-scale fading, this result reveals that the outage probability performance becomes much worse when the obstacles are in large number. For FSO systems, this result reveals that a longer link distance, which results in a larger Rytov variance, will cause significant performance degradation. We comment that it requires tens of hours to obtain the simulated results while calculating the asymptotic outage probabilities costs less than 0.1 second.

\begin{figure}
  \centering
  \includegraphics[width=0.5\textwidth]{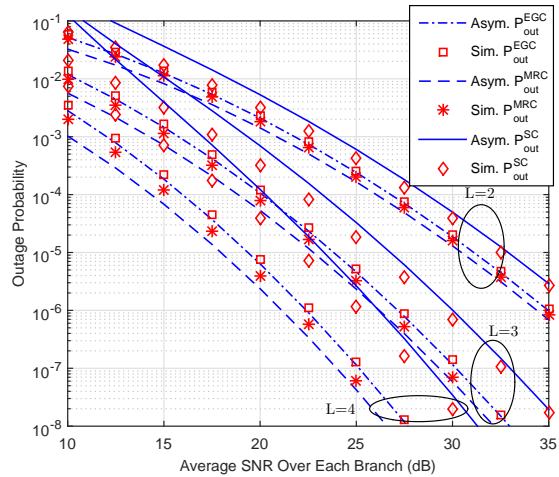}\\
  \caption{Outage probabilities of dual-branch SC, EGC and MRC over correlated lognormal channels. The parameters are $\sigma_G=1.2$, $\gamma_{th}=0.1$W, $\rho=0.1$ and $L=2,3,4$.}\label{Multi-hop}
\end{figure}
Figure \ref{Multi-hop} compares the outage probabilities of SC, EGC and MRC over lognormal fading channels with different branch number. It can be observed again that the asymptotic outage probabilities well match the exact outage probabilities. Besides, Fig. \ref{Multi-hop} shows that the performance gaps for different diversity schemes increase with the branch number. For example, the performance gap between SC and MRC is $5$dB for $L=4$ when the outage probability is $10^{-7}$, and the performance gap decreases to $3.5$dB for $L=3$. Besides, the performance gap between EGC and MRC is quite close for different $L$ values, and this is because ${{\textbf{x}}_{nst}^{EGC}} ={{\textbf{x}}_{nst}^{MRC}}$, which determines the dominant term of the integral of the outage probability.

\begin{figure}
  \centering
  \includegraphics[width=0.5\textwidth]{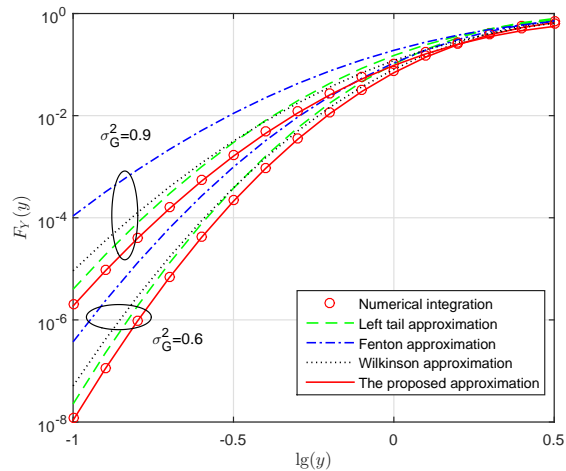}\\
  \caption{CDF of $Y=\exp(G_1)+\exp(G_2)$ and its approximations, where $G_1$ and $G_2$ are independent Gaussian RVs that have identical mean $\mu$ and variance $\sigma_G^2$. The parameters are $\mu=0$, $\sigma_G^2=0.3,0.6$. The exact CDF of sum of lognormal RVs is obtained using numerical integration. Wilkinson, Fenton and left tail approximations are, respectively, based on \cite{schwartz1982distribution},\cite{fenton1960sum} and \cite{szyszkowicz2007tails}.} \label{CompareSumOfLN}
\end{figure}
By observing \eqref{II-6.2}, it can be observed that the asymptotic outage probability for EGC is essentially the CDF of sum of lognormal RVs at $\sqrt{L\gamma_{th}}$. Therefore, we can reform \eqref{III-49} and obtain the CDF of sum of correlated lognormal RVs as
\begin{equation}\label{N-1}
\begin{split}
  F_Y(y)&=\Pr \left\{ {\sum\limits_{l = 1}^L {\exp \left( {{G_l}} \right)}  \le y} \right\} \approx 1 - {Q_{\frac{L}{2}}}\left( {\sqrt L \left( {\frac{{\frac{{\ln L + {\mu _G} - \ln y}}{{a + L - 1}} + \frac{{L - 1 + a}}{{{{\left( {1 - a} \right)}^2}}}}}{{\frac{{{\sigma _G}}}{{\sqrt {{a^2} + L - 1} }}}}} \right),\frac{{\left( {L - 1 + a} \right)\sqrt L }}{{{{\left( {1 - a} \right)}^2}\frac{{{\sigma _G}}}{{\sqrt {{a^2} + L - 1} }}}}} \right).
  \end{split}
\end{equation}
Similarly, based on \eqref{III-49.5}, we obtain the CDF of sum of independent lognormal RVs as
\begin{equation}\label{N-2}
\begin{split}
  F_Y(y)\approx 1 - {Q_{\frac{L}{2}}}\left( {\frac{{\sqrt L }}{{{\sigma _G}}}\left( {\ln L + {\mu _G} - \ln y + 1} \right),\frac{{\sqrt L }}{{{\sigma _G}}}} \right).
  \end{split}
\end{equation}
Figure \ref{CompareSumOfLN} compares the tightness of \eqref{N-2} with some existing approximations on sum of lognormal RVs. All of these considered approximations are based on closed-form formulas and can provide results in one millisecond. It can be observed that Wilkinson, Fenton and left tail approximations cannot accurately evaluate the left tail of the CDF. In contrast, the proposed approximation scheme is accurate in a large range of argument.

\section{Conclusions}\label{conclusions}
Asymptotic outage probability expressions of SC, EGC and MRC over equally correlated lognormal fading channels were derived. A theorem was developed to greatly simplify the performance analyses. Based on the derived closed-form expressions, we revealed some unique properties of diversity receptions over correlated lognormal fading channels. For example, the decreasing speed of outage probabilities was proved to be a function of the correlation coefficients; the outage probability ratio between SC and EGC (or MRC) becomes infinitely large in high SNR region. More importantly, the derivation process and the asymptotic expressions reveal insights into this long-standing problem, making it possible to evaluate outage probability of the diversity reception systems without resorting to time-consuming Monte Carlo simulation or multi-fold numerical integration.

\begin{appendices}
\section{}\label{Lemma}
This section proves the lemma in \eqref{III-2}.

\begin{proof}
Noting that $ \left| \textbf{x} - \boldsymbol{\mu} \right|/{\sigma}$ follows chi distribution whose CDF can be expressed using incomplete Gamma function, we obtain
\begin{equation}\label{III-3}
\begin{split}
&\int\limits_{{\Omega _1(\boldsymbol{\mu},\textbf{x}_0)}} {{f_{{\bf{x}},iid}}\left( {\textbf{x}} \right)d} {\textbf{x}}=\int\limits_{{{{\left| \textbf{x} - \boldsymbol{\mu} \right|}}}  > \left| {{{\textbf{x}}_0} - \boldsymbol{\mu}} \right| + \sqrt L \varepsilon+\varepsilon} {\frac{1}{{\sqrt {{{\left( {2\pi } \right)}^L}{\sigma ^{2L}}} }}\exp \left( { - \frac{1}{{2{\sigma ^2}}}{{\left|{{\textbf{x}} - {\boldsymbol{\mu}}} \right|^2}}} \right)d} {\textbf{x}}\\
&{\rm{ = }}\frac{1}{{\Gamma \left( {L/2} \right)}}\Gamma \left( {\frac{L}{2},\frac{{ {{{\left(\left| {{{\textbf{x}}_0} - \boldsymbol{\mu}} \right| + \sqrt L \varepsilon+\varepsilon\right)}^2}}}}{{2{\sigma ^2}}}} \right)
\end{split}
\end{equation}
where $\Gamma \left(\cdot,\cdot\right)$ is the incomplete gamma function defined as $\Gamma(s,x)=\int_x^\infty t^{s-1}e^{-t}dt$. Based on \cite[eq. (8.357)]{integraltable}, we expand the incomplete gamma function in \eqref{III-3} and obtain
\begin{equation}\label{III-4}
\begin{split}
  &\int\limits_{{\Omega _1}(\boldsymbol{\mu},\textbf{x}_0)} {{f_{{\bf{x}},iid}}\left( {\textbf{x}} \right)d} {\textbf{x}}{\rm{ = }}\frac{1}{{\Gamma \left( {L/2} \right)}}{\left( {\frac{{ {{{\left(\left| {{{\textbf{x}}_0} - \boldsymbol{\mu}} \right| + \sqrt L \varepsilon+\varepsilon\right)}^2}}}}{{2{\sigma ^2}}}} \right)^{\frac{L}{2} - 1}} \exp\left({ - \frac{{{{{\left(\left| {{{\textbf{x}}_0} - \boldsymbol{\mu}} \right| + \sqrt L \varepsilon+\varepsilon\right)}^2}}}}{{2{\sigma ^2}}}}\right)\\ &\times\left[ {\sum\limits_{m = 0}^{M - 1} {\frac{{{{\left( { - 1} \right)}^m}\Gamma \left( {1 - \frac{L}{2} + m} \right)}}{{{{\left( {\frac{{ {{{\left(\left| {{{\textbf{x}}_0} - \boldsymbol{\mu}} \right| + \sqrt L \varepsilon+\varepsilon\right)}^2}}}}{{2{\sigma ^2}}}} \right)}^m}\Gamma \left( {1 - \frac{L}{2}} \right)}} + O\left( {{{\left| {\frac{{ {{{\left(\left| {{{\textbf{x}}_0} - \boldsymbol{\mu}} \right| + \sqrt L \varepsilon+\varepsilon\right)}^2}}}}{{2{\sigma ^2}}}} \right|}^{ - M}}} \right)} } \right]
  \end{split}
\end{equation}
where $O(\cdot)$ is the big $O$ notation; for two functions $f(x)$ and $g(x)$, if $\lim\limits_{x\to a} \sup \left|\frac{f(x)}{g(x)}\right|<\infty$, then $f(x)=O(g(x))$.

For the integral in the denominator of \eqref{III-2}, we have
\begin{equation}\label{III-5}
\begin{split}
&\int\limits_{{\omega _2(\textbf{x}_0)}} {{f_{{\textbf{x}},iid}}\left( {\textbf{x}} \right)d} {\textbf{x}}{\rm{ = }}\int\limits_{\left| {{x_l} - {x_{l,0}}} \right| \le \varepsilon ,\forall l=1\cdots L} {\frac{1}{{\sqrt {{{\left( {2\pi } \right)}^L}{\sigma ^{2L}}} }}\exp \left( { - \frac{1}{{2{\sigma ^2}}}{{\left| {{\textbf{x}} - \boldsymbol{\mu}} \right|}^2}} \right)d} {\textbf{x}}.
\end{split}
\end{equation}
In the region ${\left| {{x_l} - {x_{l,0}}} \right| \le \varepsilon ,\forall l=1\cdots L}$, we can obtain $\sum\limits_{l = 1}^L {{{\left| {{x_l} - {x_{l,0}}} \right|}^2}}  \le L{\varepsilon ^2}$, thus
\begin{equation}\label{III-5.3}
  \left| {{\textbf{x}} - {{\textbf{x}}_0}} \right| \le \sqrt L \varepsilon.
\end{equation}
By applying the triangle inequality, we obtain
\begin{equation}\label{III-5.4}
 \left| {{\textbf{x}} - \boldsymbol{\mu}} \right| = \left| {{\textbf{x}} - {{\textbf{x}}_0} + {{\textbf{x}}_0} - \boldsymbol{\mu}} \right| \le \left| {{\textbf{x}} - {{\textbf{x}}_0}} \right| + \left| {{{\textbf{x}}_0} - \boldsymbol{\mu}} \right| \le \left| {{{\textbf{x}}_0} - \boldsymbol{\mu}} \right| + \sqrt L \varepsilon
\end{equation}
in the region ${\left| {{x_l} - {x_{l,0}}} \right| \le \varepsilon ,\forall l=1\cdots L}$, where the last inequality is based on \eqref{III-5.3}. Therefore, we can substitute \eqref{III-5.4} into \eqref{III-5} and obtain
\begin{equation}\label{III-5.5}
\begin{split}
&\int\limits_{{\omega _2(\textbf{x}_0)}} {{f_{{\textbf{x}},iid}}\left( {\textbf{x}} \right)d} {\textbf{x}}>\int\limits_{\left| {{x_l} - {x_{l,0}}} \right| \le \varepsilon ,\forall l=1\cdots L} {\frac{1}{{\sqrt {{{\left( {2\pi } \right)}^L}{\sigma ^{2L}}} }}\exp \left( { - \frac{1}{{2{\sigma ^2}}}{{\left(\left| {{{\textbf{x}}_0} - \boldsymbol{\mu}} \right| + \sqrt L \varepsilon \right)}^2}} \right)d} {\textbf{x}}\\
&= \frac{1}{{\sqrt {{{\left( {2\pi } \right)}^L}{\sigma ^{2L}}} }}{ (2\varepsilon)^L}\exp \left( { - \frac{1}{{2{\sigma ^2}}} {{{{(\left| {{{\textbf{x}}_0} - \boldsymbol{\mu}} \right|+\sqrt{L}\varepsilon)^2}}}} } \right)
\end{split}
\end{equation}
where the factor ``$(2\varepsilon)^L$'' in the last equality is the volume of the integral region. The ratio between \eqref{III-4} and \eqref{III-5.5} satisfies
\begin{equation}\label{III-9}
\begin{split}
  &\frac{{\int\limits_{{\Omega _1(\boldsymbol{\mu},\textbf{x}_0)}} {{f_{{\bf{x}},iid}}\left( {\textbf{x}} \right)d} {\textbf{x}}}}{{\int\limits_{{\omega _2(\textbf{x}_0)}} {{f_{{\textbf{x}},iid}}\left( {\textbf{x}} \right)d{\textbf{x}}} }} < \frac{{\frac{1}{{\Gamma \left( {L/2} \right)}}{{\left( {\frac{{ {{{\left(\left| {{{\textbf{x}}_0} - \boldsymbol{\mu}} \right| + \sqrt L \varepsilon+\varepsilon\right)}^2}}}}{{2{\sigma ^2}}}} \right)}^{\frac{L}{2} - 1}}{\exp{\left( - \frac{1}{{2{\sigma ^2}}}{ {{{\left(\left| {{{\textbf{x}}_0} - \boldsymbol{\mu}} \right| + \sqrt L \varepsilon+\varepsilon\right)}^2}}}\right)}}}}{{\frac{1}{{\sqrt {{{\left( {2\pi } \right)}^L}{\sigma ^{2L}}} }}{(2\varepsilon) ^L}\exp \left( { - \frac{1}{{2{\sigma ^2}}}(\left| {{{\textbf{x}}_0} - \boldsymbol{\mu}} \right|+\sqrt L \varepsilon)^2 } \right)}}\\ &\times\left[ {\sum\limits_{m = 0}^{M - 1} {\frac{{{{\left( { - 1} \right)}^m}\Gamma \left( {1 - \frac{L}{2} + m} \right)}}{{{{\left( {\frac{{ {{{\left(\left| {{{\textbf{x}}_0} - \boldsymbol{\mu}} \right| + \sqrt L \varepsilon+\varepsilon\right)}^2}}}}{{2{\sigma ^2}}}} \right)}^m}\Gamma \left( {1 - \frac{L}{2}} \right)}} + O\left( {{{\left| {\frac{{ {{{\left(\left| {{{\textbf{x}}_0} - \boldsymbol{\mu}} \right| + \sqrt L \varepsilon+\varepsilon\right)}^2}}}}{{2{\sigma ^2}}}} \right|}^{ - M}}} \right)} } \right]\\
  &=\Psi \left({{{\left| {{{\textbf{x}}_0} - \boldsymbol{\mu}} \right|}}}\right)\exp \left( { - \frac{1}{{2{\sigma ^2}}}\left( {{\varepsilon ^2}{\rm{ + }}2 \varepsilon  (\left| {{{\textbf{x}}_0} - \boldsymbol{\mu}} \right|+\sqrt{L}\varepsilon  ) } \right)} \right)
  \end{split}
\end{equation}
where $\Psi\left(x\right)$ is a polynomial of $x$ of finite order. As $|\boldsymbol{\mu}|\to\infty$, $ \left| {{{\textbf{x}}_0} - \boldsymbol{\mu}} \right|>|\boldsymbol{\mu}|-|\textbf{x}_0|\to\infty$, and the exponential term is a high order infinitely small quantity, and its order is higher than any polynomial, thus we obtain \eqref{III-2}.
\end{proof}

\section{}\label{Theorem}
This is the proof to the theorem in \eqref{III-12}.

\begin{proof}
Since  $\omega _1(\boldsymbol{\mu},\textbf{x}_0)\subset \bar \Theta \left( {{{\textbf{x}}_0}},\boldsymbol{\mu} \right)$, we obtain
 \begin{equation}\label{III-15}
   \omega _1(\boldsymbol{\mu},\textbf{x}_0)\subset \Omega_1(\boldsymbol{\mu},\textbf{x}_0)= \left\{ {{\textbf{x}}\left| {{{{\left| \textbf{x} - \boldsymbol{\mu} \right|}}}  > \left| {{{\textbf{x}}_0} - \boldsymbol{\mu}} \right| + \sqrt L \varepsilon+\varepsilon} \right.} \right\}
 \end{equation}
as long as $\varepsilon$ is sufficiently small. Therefore, we can obtain
\begin{equation}\label{III-16}
  \int\limits_{{\omega _1(\boldsymbol{\mu},\textbf{x}_0)}} {{f_{{\textbf{x}},iid}}\left( {\textbf{x}} \right)d} {\textbf{x}} < \int\limits_{{\Omega _1(\boldsymbol{\mu},\textbf{x}_0)}} {{f_{{\textbf{x}},iid}}\left( {\textbf{x}} \right)d} {\textbf{x}}
\end{equation}
for small $\varepsilon$. For $\textbf{x}_1\in\Omega _2(\textbf{x}_0, \boldsymbol{\mu})$, we obtain
\begin{equation}\label{III-16.5}
\omega_2(\textbf{x}_1)\subset\Omega _2(\textbf{x}_0, \boldsymbol{\mu})
\end{equation}
  if $\varepsilon$ is sufficiently small, where $\omega_2(\textbf{x}_1)$ is an arbitrarily small region defined in the lemma in \eqref{III-2}. According to ${{\textbf{x}}_1} \in {\Omega _2}\left( {{{\textbf{x}}_0}} \right) \subset \Theta \left( \textbf{x}_0, \boldsymbol{\mu} \right)$ and the definition of
 $\Theta \left({{{\textbf{x}}_0}}, \boldsymbol{\mu}\right)$, we obtain $ \left| {{{\textbf{x}}_1} - \boldsymbol{\mu}} \right| < \left| {{{\textbf{x}}_0} - \boldsymbol{\mu}} \right|$ which leads to
\begin{equation}\label{III-17.5}
  {f_{{\textbf{x}},iid}}\left( {{{\textbf{x}}_0}} \right) < {f_{{\textbf{x}},iid}}\left( {{{\textbf{x}}_1}} \right).
\end{equation}
Furthermore, noting that $\omega_2(\textbf{x}_0)$ and $\omega_2(\textbf{x}_1)$ are small regions when $\varepsilon$ is small, we can expand ${f_{{\textbf{x}},iid}}\left( {{{\textbf{x}}}} \right)$  using Taylor series and obtain
\begin{equation}\label{III-18}
\begin{split}
&\int\limits_{{\omega _2}\left( {{{\textbf{x}}_0}} \right)} {{f_{{\textbf{x}},iid}}\left( {\textbf{x}} \right)d} {\textbf{x}}=\int\limits_{{\omega _2\left( {{{\textbf{x}}_0}} \right)}} {{f_{{\textbf{x}},iid}}\left( {{{\textbf{x}}_0}} \right) + o\left( {{f_{{\textbf{x}},iid}}\left( {{{\textbf{x}}_0}} \right)} \right)d} {\bf{x}}={\left( {{\rm{2}}\varepsilon } \right)^K}{f_{{\textbf{x}},iid}}\left( {{{\textbf{x}}_0}} \right) + o\left( {{\varepsilon ^K}} \right)
\end{split}
\end{equation}
and similarly
\begin{equation}\label{III-19}
\begin{split}
&\int\limits_{{\omega _2}\left( {{{\textbf{x}}_1}} \right)} {{f_{{\textbf{x}},iid}}\left( {\textbf{x}} \right)d} {\textbf{x}}={\left( {{\rm{2}}\varepsilon } \right)^K}{f_{{\textbf{x}},iid}}\left( {{{\textbf{x}}_1}} \right) + o\left( {{\varepsilon ^K}} \right).
\end{split}
\end{equation}
According to \eqref{III-17.5}, \eqref{III-18} and \eqref{III-19}, we obtain
\begin{equation}\label{III-20}
  \int\limits_{{\omega _2}\left( {{{\textbf{x}}_1}} \right)} {{f_{{\textbf{x}},iid}}\left( {\textbf{x}} \right)d} {\textbf{x}}>\int\limits_{{\omega _2}\left( {{{\textbf{x}}_0}} \right)} {{f_{{\textbf{x}},iid}}\left( {\textbf{x}} \right)d} {\textbf{x}}.
\end{equation}
Based on \eqref{III-16.5} and \eqref{III-20}, we obtain
\begin{equation}\label{III-22}
  \int\limits_{{\Omega _2(\textbf{x}_0)}} {{f_{{\textbf{x}},iid}}\left( {\textbf{x}} \right)d} {\textbf{x}}>\int\limits_{{\omega _2}\left( {{{\textbf{x}}_0}} \right)} {{f_{{\textbf{x}},iid}}\left( {\textbf{x}} \right)d} {\textbf{x}}.
\end{equation}
According to \eqref{III-16}, \eqref{III-22} and the lemma in \eqref{III-2}, we obtain \eqref{III-12}.
\end{proof}

\section{}\label{KKT}
This section calculates the nearest point to $\boldsymbol{\mu}_X$ inside the integral region $\Phi_{SC}(\textbf{x})\leq0$.

The objective function is the square of the distance between $\textbf{x}$ and $\boldsymbol{\mu}_X$, which is shown as
\begin{equation}\label{A-5}
  d^2(\textbf{x},\boldsymbol{\mu}_X)=\sum\limits_{l = 1}^L {{{\left( {{x_l} - {\mu _X}} \right)}^2}}
\end{equation}
and the constraint is the integral region in \eqref{III-25}, which can be rewritten as
\begin{equation}\label{A-6}
  \left\{ \begin{array}{l}
a{x_1} + \sum\limits_{k \ne 1} {{x_k}} - \ln \sqrt{\gamma _{th}}\le 0\\
 \vdots \\
a{x_L} + \sum\limits_{k \ne L} {{x_k}} - \ln \sqrt{\gamma _{th}}\le 0.
\end{array} \right.
\end{equation}
The KKT conditions for the minimizer is \cite{boyd2004convex}
\begin{equation}\label{A-7}
  \nabla d{\left( {\textbf{x},\boldsymbol{\mu} } \right)^2} = \sum\limits_{l = 1}^L {{\lambda _l}\nabla } \left( {a{x_l} + \sum\limits_{k \ne l} {{x_k}}  - \ln \sqrt{\gamma _{th}}} \right)
\end{equation}
where $\lambda_l$s are the KKT multipliers. Equation \eqref{A-7} can be simplified to
\begin{equation}\label{A-8}
  2\left( {{x_l} - {\mu _X}} \right) = {\lambda _l}a + {\lambda _l}\left( {L - 1} \right),\forall l = 1, \cdots ,L.
\end{equation}
Assuming the $L$ constraints in \eqref{A-6} are all active, the ``$\leq$'' signs in \eqref{A-6} become ``$=$''. Together with the $L$ equations in \eqref{A-8}, we can find the minimizers as
\begin{equation}\label{A-9}
  \textbf{x}_{nst}^{SC}=[{\frac{{\ln {\gamma _{th}}}}{2({a + L - 1})} },\cdots,{\frac{{\ln {\gamma _{th}}}}{2({a + L - 1})} }]
\end{equation}
and $\lambda_l$s as $  {\lambda _1} = \lambda _2=\cdots=\lambda _L= \frac{{2\left( {{x_{nst}} - {\mu _X}} \right)}}{{a + L - 1}}$.

\section{}\label{Subsets}
This is a proof to \eqref{III-27}.

\begin{proof}
For two integral regions ${\Phi _1}{\rm{ = }}\left( {\left. {\textbf{x}} \right|f\left( {\textbf{x}} \right) < 0} \right)$ and ${\Phi _2}{\rm{ = }}\left( {\left. {\textbf{x}} \right|g\left( {\textbf{x}} \right) < 0} \right)$, if $f\left( {\textbf{x}} \right) < 0\Rightarrow g\left( {\textbf{x}} \right) < 0$, then ${\Phi _1}\subset{\Phi _2}$, where ``$\Rightarrow$'' denotes ``results in''. Therefore, we need to prove
\begin{equation}\label{A-10.5}
\left\{ \begin{array}{l}
a{x_1} + \sum\limits_{k=1,k \ne 1}^L {{x_k}}  < \ln \sqrt{\gamma _{th}}\\
 \vdots \\
a{x_L} + \sum\limits_{k=1,k \ne L}^L {{x_k}}  < \ln \sqrt{\gamma _{th}}\\
\left| {\textbf{x} - {\boldsymbol{\mu} _X}} \right| < \left| {{\textbf{x}_0} - {\boldsymbol{\mu}_X}} \right|
\end{array} \right. \Rightarrow \left\{ \begin{array}{l}
\ln \sqrt{\gamma _{th}} - L\left( {a + L - 1} \right)\varepsilon  < a{x_1} + \sum\limits_{k=1, k \ne 1}^L {{x_k}}  < \ln \sqrt{\gamma _{th}}\\
 \vdots \\
\ln \sqrt{\gamma _{th}} - L\left( {a + L - 1} \right)\varepsilon  < a{x_L} + \sum\limits_{k=1, k \ne L}^L {{x_k}}  < \ln \sqrt{\gamma _{th}}
\end{array} \right.
\end{equation}
which is rewritten from \eqref{III-27}.

The inequality $|\textbf{x}-\boldsymbol{\mu}_X|<|\textbf{x}_0-\boldsymbol{\mu}_X|$ on the left-hand side of \eqref{A-10.5} can be rewritten as
\begin{equation}\label{A-11}
  \sum\limits_{l = 1}^L {{{\left( {{x_l} - {\mu _X}} \right)}^2}}  < \sum\limits_{l = 1}^L {{{\left( {\frac{{\ln {\gamma _{th}}}}{{2(a + L - 1)}} - \varepsilon  - {\mu _X}} \right)}^2}}.
\end{equation}
Expanding the squared terms and cancelling $\mu_X^2$ on both sides, we obtain
\begin{equation}\label{A-12}
  \sum\limits_{l = 1}^L {x_l^2 - 2{x_l}{\mu _X}}  < L{\left( {\frac{{\ln {\gamma _{th}}}}{{2(a + L - 1)}}} \right)^2} - \sum\limits_{l = 1}^L {2\left( {\frac{{\ln {\gamma _{th}}}}{2({a + L - 1})} - \varepsilon } \right){\mu _X}}.
\end{equation}
By discarding $x_l^2$ on the left-hand side of \eqref{A-12}, we obtain
\begin{equation}\label{A-13}
  \sum\limits_{l = 1}^L { - 2{x_l}{\mu _X}}  < L{\left( {\frac{{\ln {\gamma _{th}}}}{2({a + L - 1})}} \right)^2} -  {2L\left( {\frac{{\ln {\gamma _{th}}}}{2({a + L - 1})} - \varepsilon } \right){\mu _X}}.
\end{equation}
As $\mu_X \to \infty$, eq. \eqref{A-13} leads to
\begin{equation}\label{A-15}
  \sum\limits_{l = 1}^L {{x_l}}  > L {\left( {\frac{{\ln {\gamma _{th}}}}{2(a + L - 1)} - \varepsilon } \right)} .
\end{equation}
Noting that \eqref{A-15} was derived from $|\textbf{x}-\boldsymbol{\mu}_X|<|\textbf{x}_0-\boldsymbol{\mu}_X|$, we can combine \eqref{A-15} with the first $L-1$ inequalities on the left-hand side of \eqref{A-10.5} and obtain
\begin{equation}\label{A-16}
  \left\{ \begin{array}{l}
a{x_1} + \sum\limits_{k=1,k \ne 1}^L {{x_k}}  < \ln \sqrt{\gamma _{th}}\\
 \vdots \\
a{x_{L - 1}} + \sum\limits_{k=1,k \ne L - 1}^L {{x_k}}  < \ln \sqrt{\gamma _{th}}\\
\sum\limits_{l = 1}^L {{x_l}}  > {L\left( {\frac{{\ln {\gamma _{th}}}}{2({a + L - 1})} - \varepsilon } \right)}.
\end{array} \right.
\end{equation}
Summing the first $L-1$ inequalities in \eqref{A-16} and multiplying both sides by $-1$, we obtain
\begin{equation}\label{A-17}
  \left\{ \begin{array}{l}
- \left( {a + L - 2} \right)\sum\limits_{l = 1}^{L - 1} {{x_l}}  - \left( {L - 1} \right){x_L} >  -  \frac{L - 1}{2} \ln {\gamma _{th}}\\
\sum\limits_{l = 1}^L {{x_l}}  > L{\left( {\frac{{\ln {\gamma _{th}}}}{2({a + L - 1})} - \varepsilon } \right)}.
\end{array} \right.
\end{equation}
Multiplying the two sides of the first inequality in \eqref{A-17} by ${(a+L-1)}^{-1}$ and adding the resulted inequality to the second inequality in \eqref{A-17}, after simplification, we obtain
\begin{equation}\label{A-18}
  \sum\limits_{l = 1}^{L - 1} {{x_l}}  + a{x_L} > \ln \sqrt{\gamma _{th}} - L\left( {a + L - 1} \right)\varepsilon .
\end{equation}
Noting that \eqref{A-16} is symmetrical for $x_l$'s, with similar procedures from \eqref{A-16} to \eqref{A-17}, we can obtain
\begin{equation}\label{A-19}
  \sum\limits_{k=1, k \neq l}^{L} {{x_k}}  + a{x_l} > \ln \sqrt{\gamma _{th}} - L\left( {a + L - 1} \right)\varepsilon, \forall l=1,\cdots,L .
\end{equation}
Combining \eqref{A-19} and the first $L$ inequalities in the left-hand side of \eqref{A-10.5}, we obtain the right-hand side of \eqref{A-10.5}.
\end{proof}

\section{}\label{Derivatives}
In this appendix, we prove that for the two hypersurfaces $\Phi_{EGC}(\textbf{x})=0$ and $\tilde \Phi_{EGC}(\textbf{x})=0$, the first-order derivatives ${\frac{{\partial {x_1}}}{{\partial {x_m}}}}$'s are identical at $\textbf{x}_{nst}$, and the second-order derivatives ${\frac{{\partial x_1^2}}{{\partial {x_m}\partial {x_n}}}}$ are also identical at $\textbf{x}_{nst}$, where $x_1$ is regarded as a function of $x_m$'s and $m,n=2,\cdots,L$.

\begin{proof}
By taking the partial derivatives of both sides of
\begin{equation}\label{A-26}
  \Phi_{EGC}(\textbf{x})=\sum\limits_{l = 1}^L {\exp \left( {a{x_l} + \sum\limits_{k = 1,l \ne k}^L {{x_k}} } \right)}  - \sqrt {L{\gamma _{th}}}=0
\end{equation}
in terms of $x_m$, we can obtain
%\begin{equation}\label{A-27}
%  \begin{split}
%&\exp \left( {a{x_1} + \sum\limits_{k = 2}^L {{x_k}} } \right)\left( {a\frac{{\partial {x_1}}}{{\partial {x_m}}} + \sum\limits_{k = 2}^L {\frac{{\partial {x_k}}}{{\partial {x_m}}}} } \right) \\&+ \sum\limits_{l = 2,l \ne m}^L {\exp \left( {a{x_l} + \sum\limits_{k = 1,k \ne l}^L {{x_k}} } \right)\left( {a\frac{{\partial {x_l}}}{{\partial {x_m}}} + \sum\limits_{k = 1,k \ne l}^L {\frac{{\partial {x_k}}}{{\partial {x_m}}}} } \right)} \\
% &+ \exp \left( {a{x_m} + \sum\limits_{k = 1,k \ne m}^L {{x_k}} } \right)\left( {a\frac{{\partial {x_m}}}{{\partial {x_m}}} + \sum\limits_{k = 1,k \ne m}^L {\frac{{\partial {x_k}}}{{\partial {x_m}}}} } \right) = 0
%\end{split}
%\end{equation}
%which can be simplified to
\begin{equation}\label{A-28}
  \begin{split}
\exp \left( {a{x_1} + \sum\limits_{k = 2}^L {{x_k}} } \right)\left( {a\frac{{\partial {x_1}}}{{\partial {x_m}}} + 1} \right) + \sum\limits_{l = 2,m \ne l}^L {\exp \left( {a{x_l} + \sum\limits_{k = 1,l \ne k}^L {{x_k}} } \right)\left( {\frac{{\partial {x_1}}}{{\partial {x_m}}} + 1} \right)} \\
 + \exp \left( {a{x_m} + \sum\limits_{k = 1,k \ne m}^L {{x_k}} } \right)\left( {a + \frac{{\partial {x_1}}}{{\partial {x_m}}}} \right) = 0.
\end{split}
\end{equation}
Substituting $x_2=x_3=\cdots=x_L=x_{nst}^{EGC}$ into \eqref{A-28} and after some simplification, we obtain
\begin{equation}\label{A-29}
  \frac{{\partial {x_1}}}{{\partial {x_m}}}=-1.
\end{equation}

Taking partial derivatives of both sides of \eqref{A-28} in terms of $x_n$, we obtain
\begin{equation}\label{A-30}
  \begin{split}
\left[ {\exp \left( {a{x_1} + \sum\limits_{k = 2}^L {{x_k}} } \right)\left( {a\frac{{\partial {x_1}}}{{\partial {x_n}}} + 1} \right)\left( {a\frac{{\partial {x_1}}}{{\partial {x_m}}} + 1} \right) + \exp \left( {a{x_1} + \sum\limits_{k = 2}^L {{x_k}} } \right)a\frac{{{\partial ^2}{x_1}}}{{\partial {x_m}\partial {x_n}}}} \right]\\
 + \sum\limits_{l = 2,m \ne l}^L {\left[ {\frac{\partial }{{\partial {x_n}}}\left( {\exp \left( {a{x_l} + \sum\limits_{k = 1,l \ne k}^L {{x_k}} } \right)} \right)\left( {\frac{{\partial {x_1}}}{{\partial {x_m}}} + 1} \right) + \exp \left( {a{x_l} + \sum\limits_{k = 1,l \ne k}^L {{x_k}} } \right)\frac{{\partial^2 {x_1}}}{{\partial {x_m}\partial {x_n}}}} \right]} \\
 + \left[ {\frac{\partial }{{\partial {x_n}}}\left( {\exp \left( {a{x_m} + \sum\limits_{k = 1,k \ne m}^L {{x_k}} } \right)} \right)\left( {a + \frac{{\partial {x_1}}}{{\partial {x_m}}}} \right) + \exp \left( {a{x_m} + \sum\limits_{k = 1,k \ne m}^L {{x_k}} } \right)\frac{{\partial^2 x_1}}{{\partial {x_m}\partial {x_n}}}} \right] = 0.
\end{split}
\end{equation}
Substituting $m=n$, $x_2=x_3=\cdots=x_L=x_{nst}^{EGC}$ and $\frac{{\partial {x_1}}}{{\partial {x_m}}}=\frac{{\partial {x_1}}}{{\partial {x_n}}}=-1$ into \eqref{A-30}, we obtain
\begin{equation}\label{A-32}
  \frac{{{\partial ^2}{x_1}}}{{\partial {x_m^2}}} =  - \frac{{2{{\left( {a - 1} \right)}^2}}}{{\left( {L - 1 + a} \right)}}.
\end{equation}
When $m\neq n$, we substitute $x_2=x_3=\cdots=x_L=x_{nst}^{EGC}$ and $\frac{{\partial {x_1}}}{{\partial {x_m}}}=\frac{{\partial {x_1}}}{{\partial {x_n}}}=-1$ into \eqref{A-30} and obtain
\begin{equation}\label{A-34}
  \frac{{\partial^2 {x_1}}}{{\partial {x_m}\partial {x_n}}} =  - \frac{{{{\left( {1 - a} \right)}^2}}}{{\left( {L - 1 + a} \right)}}.
\end{equation}

Following a similar procedure, it is straightforward to show that for the hypersurface $\tilde \Phi_{EGC}(\textbf{x})=0$, the first and second order derivatives are $\frac{{\partial {x_1}}}{{\partial {x_m}}} =  - 1$ which are identical to the first-order derivatives in \eqref{A-29} for $\Phi_{EGC}(\textbf{x})$ and
\begin{equation}\label{A-37}
  \frac{{{\partial ^2}{x_1}}}{{\partial {x_m}\partial {x_n}}} = \left\{ {\begin{array}{*{20}{c}}
{ - \frac{{{{\left( {1 - a} \right)}^2}}}{{\left( {L - 1 + a} \right)}},m \ne n}\\
{ - \frac{{2{{\left( {1 - a} \right)}^2}}}{{\left( {L - 1 + a} \right)}},m = n}
\end{array}} \right.
\end{equation}
which are identical to the second-order derivatives in \eqref{A-32} and \eqref{A-34} for $\Phi_{EGC}(\textbf{x})=0$.
\end{proof}

\end{appendices}

\bibliographystyle{IEEEtran}

\end{document}